\documentclass[twocolumn]{aastex61}
\usepackage{graphicx}
        
        \renewcommand{\columnsep}{0.5cm}

        \renewcommand{\baselinestretch}{0.97}

\def\arcmin{\hbox{$^\prime$}}
\def\arcsec{\hbox{$^{\prime\prime}$}}

\def \eg{{\em e.g.}}
\def \gtw{\>\hbox{\lower.25em\hbox{$\buildrel >\over\sim$}}\>}
\def \ltw{\>\hbox{\lower.25em\hbox{$\buildrel <\over\sim$}}\>}

\def \ie{{\em i.e.}}

\def \radm2{rad/m$^2$}

\newcommand\ffh{\mbox{$ \!\!^{\mathrm h}$}}%
\newcommand\ffm{\mbox{$ \!\!^{\mathrm m}$}}%
\newcommand\ffs{\mbox{$ \!\!^{\mathrm s}$}}%

\def\gs{\mathrel{\raise0.35ex\hbox{$\scriptstyle >$}\kern-0.6em
\lower0.40ex\hbox{{$\scriptstyle \sim$}}}}
\def\ls{\mathrel{\raise0.35ex\hbox{$\scriptstyle <$}\kern-0.6em
\lower0.40ex\hbox{{$\scriptstyle \sim$}}}}

\shortauthors{Owen}
\shorttitle{Deep VLA Imaging of GOODS-N}

\begin{document}

\title{Deep JVLA Imaging of GOODS-N at 20cm}

\author{Frazer\,N.\ Owen}
\affiliation{National Radio Astronomy 
Observatory, P.\ O.\ Box O,
Socorro, NM 87801 USA\footnote{The National Radio Astronomy
Observatory is facility of the National Science Foundation operated
under cooperative agreement by Associated Universities Inc.}}

\correspondingauthor{Frazer\,N.\ Owen\ }
\email{fowen@nrao.edu}

\begin{abstract}

New wideband continuum observations in the $1-2$ GHz band of the
GOODS-N field using NSF's Karl G. Jansky Very Large Array (VLA) 
\footnote{Since this paper compares surveys before and after the
  upgrade of the VLA electronics and its renaming, I will use ``JVLA''
for the VLA name for these new observations when the distinction
between the old and new VLA is important.}
 are presented. The best image with an
effective frequency of 1525 MHz reaches an {\it rms} noise in the
field center of $2.2\mu$Jy with $1.6$\arcsec\ resolution. A catalog of
795 sources is presented covering a radius of nine arcminutes centered
near the nominal center for the GOODS-N field, very near the nominal
VLA pointing center for the observations. 
Optical/NIR identifications and redshift estimates both from
ground-based and HST observations are discussed.  Using these
  optical/NIR data,  it is most likely that fewer than 2\% of
  the sources without confusion problems do not have a correct
  identification. A large subset of the detected sources have radio
  sizes $> 1$\arcsec. It is shown that the radio orientations for such
  sources correlate well with the HST source orientations especially
  for $z< 1$. This suggests that a least a large subset of the
  10kpc-scale disks of LIRG/ULIRG galaxies have strong star-formation,
  not just in the nucleus. For the half of the objects with $z>1$, the
  sample must be some mixture of very high star-formation rates,
  typically 300 M$\odot$/yr assuming pure star-formation, and an AGN
  or a mixed AGN/star-formation population.

\end{abstract}


\section{Introduction}

Deep radio continuum surveys combined with data from other wavelength
bands can shed new light on the nature of galaxy evolution. Both AGN
and star-forming galaxies produce synchrotron emission which can be
detected at centimeter wavelengths. The radio results can be compared
with deep surveys in other bands to determine which processes are most
important for a given object. Furthermore the surface brightness
distribution in the radio can shed light on the detailed physics. 
   AGN activity is usually associated with jets, lobes, and/or
  extremely compact, usually self-absorbed emission. Star-formation
  always has brightness temperatures $<10^5 K$ and can either be due
  to a relatively compact
  nuclear star-burst or much larger star-formation distributed within
  a galaxy's disk \citep{c92}. Synchrotron emission more extended than
  an individual galaxy can also be associated with an ongoing
  galaxy-galaxy merger \citep[e.g.][]{chj02, cb04}. Very extended
  emission can also be associated with galaxy winds, often along the
  minor axis of a galaxy, as in the local system M82
  \citep{so91,a13}. Evidence for powerful galactic winds is known
  from optical line emission \citep[e.g.][]{m12} and we might
  well expect to see low surface brightness analogs in the radio
  band. Finally, AGN are known to blow bubbles in local cooling cores,
  which are evidence of feedback which suppresses cooling
  \citep[e.g.][]{O00,p13}. At higher redshifts, such core-halo radio
  emission or radio emission associated with other AGN tracers would
  be consistent with such phenomena and would be important to
  understanding the role of AGN feedback in galaxy evolution.

However, deep radio surveys with arcsecond scale resolution
  with current radio arrays are only just able to detect radio surface
  brightness which allows the detection of such extended structures
  Low S/N means that when we detect such
  structures very little detail will be revealed. Thus we need to
  survey regions of the sky which have been well studied at other
  wavelengths to give us clues as to what we are observing.
          
 GOODS-N  \citep{d03} is one of a few very well
studied extragalactic survey regions in the sky. It contains
the original Hubble deep field \citep{w96} and
has been expanded to include the best or near-best ancillary
data in every wavelength region. The field was originally
observed at 1.4 GHz with the VLA in 1996 \citep{r00} and  those
  data were combined with Merlin imaging to produce a higher
  resolution image of a smaller field \citep{m05}.
  The VLA survey was most recently improved by \citet{m10} who
  increased the total integration time to 165 hours. For VLA
observations the GOODS-N high declination (62\arcdeg) is
ideal for deep imaging since it can be observed above 20\arcdeg\ for
15 hours and the uv-tracks are almost circular producing
very good imaging. However, some bright confusing radio sources
well outside the GOODS-N field itself have limited previous images
to noise levels worse than theoretical. The upgraded JVLA now
allows much wider bandwidths to be observed simultaneously (1-2 GHz 
in the L-band) than the old VLA  surveys which typically used
a total bandwidth
of 44 MHz at 1.4 GHz (20cm). Besides improving the sensitivity per
unit time, the wider bandwidths produce more complete uv-coverage
and thus a better synthesized beam. This better beam
reduces the impact of the confusing sources on imaging and allows
a deeper survey. Other systematics which affected the earlier imaging
are also less of a problem, \ie\ the image-plane smearing due to
the integration time and the channel bandwidth. The overall effect of
the new VLA hardware is to produce a superior image with fewer
systematics and less drop-off in sensitivity with distance from the
field center.

  Several
  papers \citep{b12,t13,RO,b14,t14,two14,b15,b17,c17,l17} have made
  use of the data reported here and this paper provides the long
  promised documentation for the survey as well as the catalog of the
  sources with peak flux densities above  $5\sigma$.  
  The discussion is also intended to put the
  survey in perspective for future studies, as well as discussing some
  of the properties of the extended sources and their optical
  identifications.

\section{Observations and Reductions}

The GOODS-N field was observed in the A-configuration for a total of
39 hours including calibration and move time between August 9 and
September 11, 2011.  A single field was observed, centered at 12 \ffh
36 \ffm 49.4 \ffs, 62\arcdeg12\arcmin 58\arcsec\ .  Eight different
scheduling blocks were observed, each of 5 hours, except for one which
was 4 hours long. Roughly 33 hours of this time were spent
on-source. The observations covered the bands from $1000-1512$ and
$1520-2032$ MHz using 1 MHz channels. A phase, bandpass and
instrumental polarization calibrator, J1313+6735, was observed every
twenty minutes. 3C286 was observed to calibrate the flux density scale
and the polarization position angles.

For each scheduling block, the data were edited and calibrated in AIPS
in the standard way. The worst parts of the band, in particular
between 1520 and 1648 MHz, were flagged at the beginning of this
process.  Also about 12\% of the data were flagged due to 
 subband edge effects in the JVLA WIDAR correlator. The rest of the
 dataset was edited using the RFLAG task, 
  resulting in a few percent more of the total being removed.
  The data were generally well-behaved and required no unusual editing. A
delay correction was calculated using the task FRING. The bandpass was
then calculated without any calibration except for the delay
correction.  After total intensity calibration, the uv-data weights
were calibrated using the AIPS task, REWAY. REWAY tends to produce
  lower weights in regions with more interference and near the edges
  of the 1000-2000 GHz band, where the system is less sensitive.
Details of the
polarization calibration and results are discussed in \citet{RO}.

\section{Imaging and Self-calibration}

Narrow band images were first made with one five hour track. These
images were used as a model for self-calibration of all the
subbands. The flux density calibration for each subband was held fixed
for each track/subband during self calibration. The task EDITA was
further used to edit out uv-data which had large discrepant solutions.
 In practice $<1$\% of the data were edited out as a result of
  this step. The whole process was repeated until all the solutions
converged with small errors. The uv-data were then merged and optimally
averaged for each subband/baseline to minimize the size of the dataset
to be imaged without distorting the image, using the AIPS procedure,
STUFFR. The uv-data were then ported to CASA.

The total intensity data were imaged in CASA using the {\tt clean}
task. In particular, wide-field, nterms=2, and Briggs weighting with
robust=0.5 were used.  For these parameters, the
Multi-Scale-Multi-Frequency-Synthesis algorithm was used
\citep{u11}. This imaging algorithm solves for the total intensity and
spectral index image across the full bandwidth, in this case 1-2
GHz. The W-projection was also used which corrects for the three
dimensional sky curvature. The scales used in the imaging were
0.35\arcsec$\times(0,3,10)$, where 0.35\arcsec\ was the pixel size
used in {\tt clean}. For detailed testing of this algorithm and
comparison to other imaging approaches see \citet{u16} and references
therein.  After imaging the CASA task {\tt widebandpbcor} was used to
correct for the primary beam attenuation.  This combination of
  parameters produced an image with 1.6\arcsec\ resolution and a {\it rms}
  noise of $2.2\mu$Jy, before the primary beam correction. The
    usually assumed parameters in the VLA Exposure Calculator for 33
    hours on-source, robust weighting and 600MHz bandwidth yield an
    expected {\it rms} of $1.9\mu$Jy.

\section{Analysis Region}

\begin{figure*}[htb!]
\epsscale{0.98} \includegraphics[trim=1.0cm 4.0cm 1.0cm 4.0cm,clip=True]{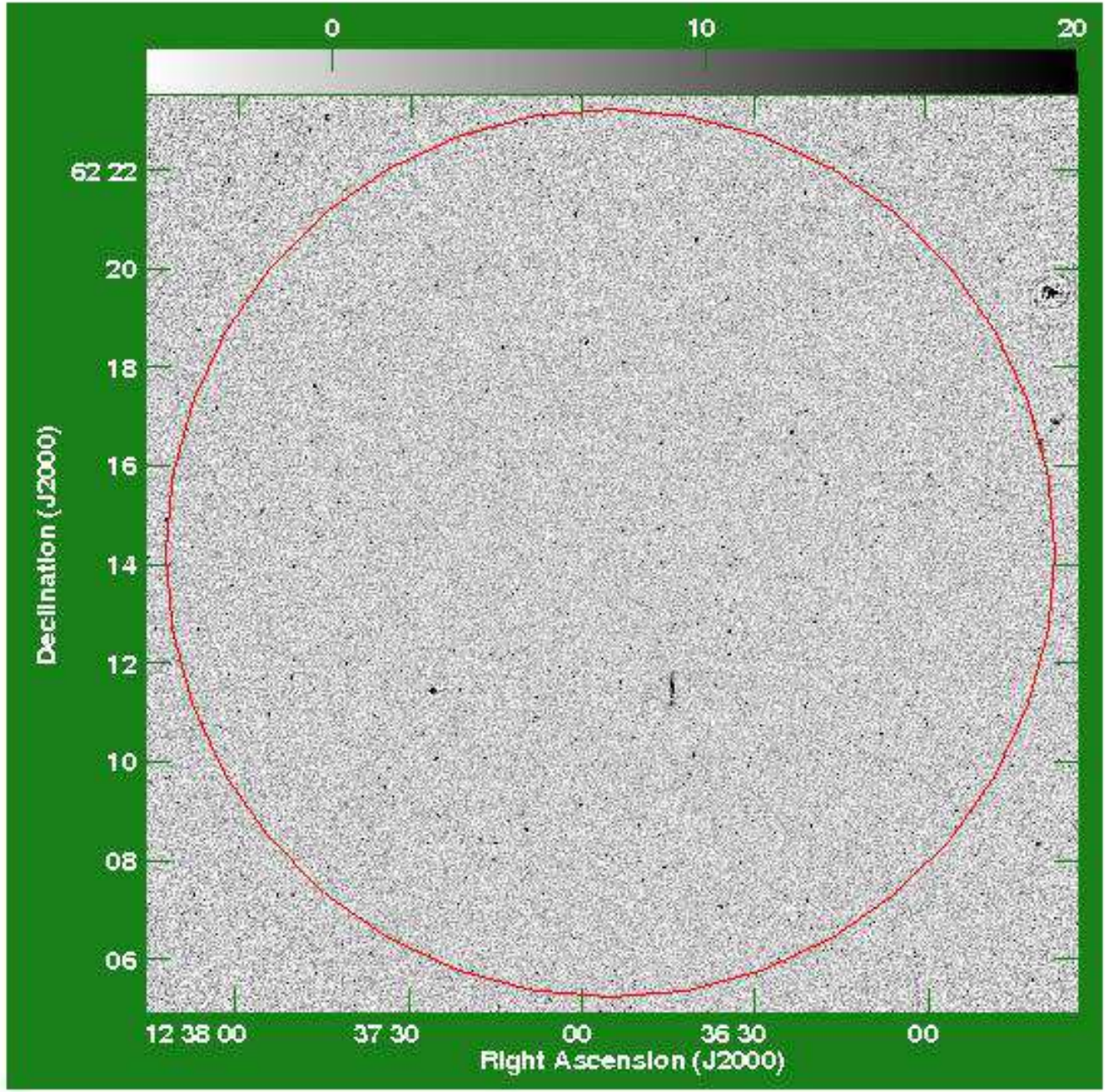}
\caption{Grey Scale 1.6\arcsec\  GOODS-N radio
  image. 1.6\arcsec$\times$1.6\arcsec\ clean beam is shown as a very small
  blue dot
  in lower left corner of the image. The red circle is the 9\arcmin\
  radius limit of the catalog. The grey scale wedge on the top of the
  image shows the range of intensities displayed from $-5\mu$Jy to $20\mu$Jy.
\label{grey}}
\end{figure*}

The  single pointing center for these VLA observations was 
12 \ffh 36 \ffm 49.4 \ffs, 62\arcdeg12\arcmin 58\arcsec\ (J2000.0), the
same as used by \citet{r00} and \citet{m10}. As such the
imaging covers a region almost 1\arcdeg\ on each side and has useful
results out to at least 20\arcmin\ from the pointing center. However,
this paper is focused on the GOODS-N area and the analysis is
only of a region of the image 9\arcmin\ in radius, centered on 
12 \ffh 36 \ffm 55.10 \ffs, 62\arcdeg14\arcmin 15\arcsec\ , which contains
all the original GOODS-N HST fields. Thus the field chosen for analysis
is about 1.5\arcmin\ from the pointing center and contains some
sources just outside the GOODS-N region. Besides limiting the analysis
to the regions studied most intensely at all wavelengths, the 9\arcmin\
radius,  centered very close to the pointing center, limits the study to
the most sensitive part of the VLA primary beam, where the smallest
instrumental corrections are required.  In figure~\ref{grey}, we show the
1.6\arcsec\ resolution image and the 9\arcmin\ survey region for reference.

\section{Results}

\subsection{The Catalog}

\subsubsection{Catalog Construction \label{cc}}

The catalog was constructed using the AIPS image cataloging task,
SAD. However, SAD, like other such cataloging tools, is best at
finding and cataloging sources with sizes near the resolution of the
image the task is given to analyze. If one wants to catalog sources
with a wider range of sizes something more should be done.

Our final images were analyzed in several different ways in order to
assemble the most complete list of sources. The main complication is
the fact that a significant fraction of the sources are resolved. The
most useful source cataloging algorithms involve fitting two
dimensional Gaussians to the significant ``islands'' of emission in
the image. That is a good technique as long as the sources are only
slightly resolved. When the true angular size of a source is bigger
than about one clean restoring beam in size, the assumption of a
Gaussian for the shape of the source in the image becomes less than
optimum. If a source is very much larger than the clean beam, many
Gaussians may be necessary to describe the emission, which are not
useful for cataloging and identifying individual sources. If a source
is bigger than the clean beam, then peak brightness may fall below the
noise cutoff and/or the fit may also not be optimum.

In order to deal with this situation, the final radio image was
convolved to several different, lower resolutions. The fitting
was done on each image and the results compared. Then the best fit for
each source was picked usually based on the resulting signal-to-noise
ratio (SNR). When a lower resolution image revealed a significantly
higher total flux density that fit was picked. Also if the fit to a
higher resolution image with a similar but slightly lower SNR more
clearly showed that a source was significantly resolved, then that fit
was chosen. Finally, for the sources which were very resolved in all
the images, the AIPS task TVSTAT, was used to estimate the total flux
density. In this case, the flux density errors were based on the total two
dimensional size, and the total size was estimated by hand, 
\ie\ by measuring X,Y positions of the source edges on the AIPS TV
  display.
   
The AIPS task SAD was used to create the first version of the source
list at each resolution. Images with circular clean beams  were
calculated from the full resolution image with sizes of 1.6\arcsec,
2.0\arcsec, 3.0\arcsec, 6.0\arcsec, and 12.0\arcsec, using the task
CONVL. For analysis of sources with peak flux densities $> 40\mu$Jy,
an extra, higher resolution image was made using robust=0 weighting in
CASA {\tt clean}. This produced an image with a clean beam of
1.02\arcsec$\times$0.82\arcsec\ at PA=82.75. When this image produced
the best fit, it is called ``1.0'' in the output table.  For each of
these images, an image of the local {\it rms} noise was calculated using the
RMSD and these images were used as input to SAD. Each image was search
down to peaks of $4.5\times$ the local rms, \ie\ $4.5\sigma$. A
collated source list from these fits was then fitted by hand using
JMFIT, which uses the same Gaussian fitting algorithm as SAD.

For the final fits sources were considered to be resolved if the total
fitted flux density exceeded the fitted peak by more than the
estimated $1\sigma$ error in the total and the minimum deconvolved,
fitted size of the major axis was $>0$. The best fit in peak $\sigma$'s
was then chosen above $5\sigma$ for each source.  If the source was
resolved at one or more resolutions, the best fit  from JMFIT or SAD
which produced an estimate of the minimum source size $>0$ was chosen.
 Note that our approach to identifying a resolved source is different
  from the $2\sigma$ criterion described by \citet{v16}, which seems
  too restrictive based on our results. Perhaps this is because the
  fits are to Gaussians in both our study and \citet{v16} and the sources
  are generally not well described as Gaussians if they have sizes
  in a cleaned image much larger than a clean beam.
  In any case, the radio-optical alignment results
  described later in this paper appear to confirm the validity of
  our method.

\begin{deluxetable*}{lrrrrrrrrrrrrrrr}[htb!]
\tabletypesize{\scriptsize}
\tablecolumns{16}
\tablewidth{0pt}
\tablenum{1}
\tablecaption{GOODS-N Radio Source Catalog \label{RC}}
\pagestyle{empty}
\tablehead{
\colhead{Name} & 
\colhead{RA} &
\colhead{err} &
\colhead{Dec} &
\colhead{err} &
\colhead{Pk} &
\colhead{err} &
\colhead{Tot} &
\colhead{err} &
\colhead{S/N} &
\colhead{$<$} &
\colhead{Ma} &
\colhead{Mi} &
\colhead{PA} &
\colhead{Res} &
\colhead{z$^a$}\\
\colhead{} &
\colhead{2000.0} &
\colhead{s} &
\colhead{2000.0} &
\colhead{\arcsec} &
\colhead{$\mu$J/b} &
\colhead{} &
\colhead{$\mu$Jy} &
\colhead{} &
\colhead{} &
\colhead{} &
\colhead{\arcsec} &
\colhead{\arcsec} &
\colhead{\arcdeg} &
\colhead{\arcsec} &
\colhead{}\\
\colhead{(1)} &
\colhead{(2)} &
\colhead{(3)} &
\colhead{(4)} &
\colhead{(5)} &
\colhead{(6)} &
\colhead{(7)} &
\colhead{(8)} &
\colhead{(9)} &
\colhead{(10)} &
\colhead{(11)} &
\colhead{(12)} &
\colhead{(13)} &
\colhead{(14)} &
\colhead{(15)} &
\colhead{(16)}}
\startdata
J123627.5+621026&12 36 27.53&0.02&62 10 26.1&0.1&17.4&2.6&17.4&2.7&6.7&$<$&1.2&\omit&\omit&1.6&0.761sB\\
J123627.5+621218&12 36 27.56&0.02&62 12 18.0&0.1&17.1&2.5&17.1&2.6&6.7&$<$&1.4&\omit&\omit&1.6&noz\\
J123627.7+621158&12 36 27.78&0.01&62 11 58.6&0.1&22.8&2.6&22.8&2.7&8.9&$<$&1.2&\omit&\omit&1.6&3.388sH\\
J123628.4+622052&12 36 28.44&0.02&62 20 52.9&0.2&15.4&3.1&15.4&3.1&5.0&$<$&0.8&\omit&\omit&1.6&1.307pQ\\
J123628.6+622139&12 36 28.67&0.01&62 21 39.7&0.1&37.3&3.2&37.3&3.4&11.5&$<$&1.2&\omit&\omit&1.6&1.871pQ\\
J123628.9+620615&12 36 28.97&0.01&62 06 15.9&0.1&45.2&2.9&57.5&6.1&15.5&\omit&1.5&0.5&90&2.0&1.264sB\\
J123629.0+621045&12 36 29.04&0.01&62 10 45.6&0.1&72.6&3.2&91.1&7.1&22.7&\omit&2.2&0.5&75&3.0&1.013sB\\
J123629.3+621613&12 36 29.36&0.08&62 16 13.9&0.5&41.6&6.2&59.3&13.9&9.0&\omit&5.0&2.6&98&6.0&0.848sB\\
J123629.3+621936&12 36 29.36&0.02&62 19 36.3&0.2&17.4&2.9&17.4&2.9&6.0&$<$&1.6&\omit&\omit&1.6&1.005sB\\
J123629.4+621513&12 36 29.45&0.02&62 15 13.2&0.2&14.2&2.5&14.2&2.5&5.6&$<$&1.8&\omit&\omit&1.6&3.652sB\\
\enddata
\tablenotetext{a}{Letter codes: s:spectroscopic, p:photometric;
  B:~\citet{b08},
  B03:~\citet{b03},
  C:~\citet{c05},
  Co:~\citet{c04},
  H:~\citet{m16},
  M:~\citet{w15}, 
  R:~\citet{r06},
  Q:~\citet{y14},
  W:~\citet{w10}}
\end{deluxetable*}

In JMFIT and SAD the minimum and maximum values were obtained by
deconvolving the source beam parameters with all 27 combinations of
the AIPS adverb EFACTOR$\times(-1, 0, 1)\times$ the uncertainties in
the major axis, minor axis, and position angle determined from the fit
added to the nominal beam size. Then the maximum and minimum values of
the deconvolved sizes and position angle were adopted as the fitted
limits on the source size.  In order to determine the best value of
EFACTOR, 10,000 sources whose peak had a range of SNR from 4 to 50 and
which were either points or one arcsecond FWHM Gaussians were
convolved with the clean beam of the 1.6\arcsec\ resolution image and
added to random points near the center of that image of the 1.6\arcsec\
resolution image. A range of values for EFACTOR were then tried for
the fit to each source in order to minimize the number of falsely
resolved sources and minimize the number of one arcsecond sources
which were found to be resolved. EFACTOR$=1.3$ produced the best
overall results. For example for S/N$=5.0$ with EFACTOR$=1.3$, 2.5\%
of source point sources were found to be resolved and 24\% of the one
arcsecond Gaussians were classified as resolved with essentially all
of the remaining sources given upper size limits $>1$\arcsec.

\subsubsection{Catalog Content}

A total of 795 radio sources with a $S/N=5$ or greater in at least one
of the images were cataloged using the procedures discussed above. In
table~\ref{RC}, we show the first ten lines of the full online table
listing the parameters of those sources.  Column 1 contains the J2000
source name. In columns $2-5$ we give the J2000 RA, the J2000
Declination and the estimated errors. Columns 6 and 7 contain the peak
flux density and the associated error.  Columns 8 and 9 list the best
estimate of the total flux density and its error. If the total fitted
flux density is more than one sigma greater than the peak and the
lower limit to the major axis size is greater than 0.0, the total
fitted flux density is quoted. Otherwise, the source is assumed to be
best represented by a point source and peak flux density and error are
quoted here. A 3\% scale error has been added in quadrature to account
for errors in the flux density calibration and the primary beam
correction. In column 10, we give the SNR of the peak detection.  If
the source is approximated as a point, there is a $<$ sign in
column 11 and the associated upper limit is given in column 12. If the
source is resolved then columns 12,13 and 14 give the best estimate of
the true source size after correcting for the clean beam. Column 15
gives the clean beam size for the image used for the quoted fit. 
  In column 16 we list the redshift. After the numeric redshift, the
  first letter indicates whether the redshift is spectroscopic (s) or
  photometric (p). The letters after that indicate the reference for
  the redshift, as detailed in the table footnotes.

Note that part of J123538.5+621643 is in the nine arcminute field of
this survey but not the parent galaxy and thus this source is
not included in the catalog (see fig~\ref{grey}).

\subsection{Identifications}

\subsubsection{NIR Identifications}

One way to understand the reliability of the catalog is to study the
identifications at other bands.  Of course, such studies also
  are necessary to understand the nature of the radio sources.  For
  this purpose the largest number of identifications are found in deep
  NIR catalogs since deep optical catalogs produce fewer
  identifications and are more confused by background sources
  \citep[\eg][]{s10,p15,v16}. This is likely because the optical
bands are biased against dusty objects and/or the objects have $z \gs 1$,
since standard optical bands are redshifted below the
4000\AA\ break. Also the SEDs of the old stellar population
stars slowly rise as one follows them from the red rest-frame optical
bands into the NIR. We have used the deep Ks catalog of \citet{w10}
  as our primary reference for identifications, searching for NIR
  identifications within a reasonable distance of the radio source
  centroid in order to be considered an identification. The
      Wang Ks catalog was obtained with using data from the
      ground-based CFHT but also includes 3.6-8.0$\mu$ flux densities
      derived from the lower spatial resolution images made with the
      Spitzer space telescope.

Given the depth of both the radio and NIR data available in
  GOODS-N, the identification problem is to find the real IDs
  superimposed on a field of random NIR sources. In every search area
  for a radio ID there will be a random source if one can search to
  deep enough levels. The search areas tend to be small, mostly much
  less than 1\arcsec\ in radius, so using data with typical
  ground-based seeing the real ID and any random IDs will produce a
  blend. With such very small search areas and NIR data generally
  deeper than is necessary for an ID, in the few cases where there is
  more than one object in the search box the brightest object is
  chosen, unlike \citet{ss92}.  Furthermore, one needs to allow for
  the possibility that the radio emitting region will not be exactly
  centered on the NIR galaxy image, either due to a non-uniform
  distribution of emission or dust obscuration even in the NIR.  The
  non-uniform sensitivity in the NIR image and the clustering of the
  NIR sources also need to be considered.  Such a search must be a
  compromise since the size of the radio source should be taken into
  account, enlarging the search area. It thus seems necessary to
  search over a larger area for the more resolved sources. However,
  the deeper the Ks catalog, the higher will be the number of random
  identifications in the larger search area. Thus to get the most
  reliable results, it is necessary to limit the search parameters,
  even if some correct identifications are missed in the process. The
  \citet{w10} catalog turns out  to be deeper than is necessary
  for the vast majority of identifications, so it makes sense to cut
  off the search at some Ks flux density in order to reduce the number
  of false identifications. However, a small subset of the radio
  sources are either at very high redshifts and/or are very red,
  perhaps due to dust extinction. These sources, while sometimes
  detected at very faint levels in the \citet{w10} Ks survey, often
  show up easily in the Spitzer IRAC bands.

To deal with these complications, we  begin by performing a search for
identifications in the \citet{w10} catalog with a restricted set of
search criteria.  In order to be considered an identification, sources
in \citet{w10} were required to be within $radius$.

\begin{equation} 
radius=(0.2^2+PE^2+SE^2)^{0.5} {\rm arcseconds} \label{eq1}
\end{equation}

where $radius$ is distance of a Ks identification from the radio
source centroid in arcseconds; $PE$ is twice the radio position error
estimate from JMFIT (columns 3 and 5 of table~\ref{RC} added in
quadrature; $SE=0.3\times Bmaj$ and $Bmaj$ is the fitted major axis
size or the upper limit to that size from column 12 of
table~\ref{RC}. $SE$ is limited to a maximum of 1.0\arcsec. Before
$radius$ was calculated the RA in \citet{w10} was increased by
0.2\arcsec\ to optimize the agreement between the two catalogs. This
offset seems to come from an offset between the reference frame in
\citet{m10} and the catalog reported here, since the \citet{w10}
coordinate system was referenced to \citet{m10} frame of reference
(see section \ref{Mcat}). The term $0.2^2$ in equation~\ref{eq1}
  is empirical, taking into account the range of offsets between the
  individually measured radio and NIR positions, and is not related to
  the mean catalog offsets.

\begin{deluxetable*}{lllrrrlll}[htb!]
\tabletypesize{\scriptsize}
\tablecolumns{9}
\tablewidth{0pt}
\tablenum{2}
\tablecaption{Sources without IDs based on equation~\ref{eq1}. \label{NID}}
\pagestyle{empty}
\tablehead{
\colhead{Name} & 
\colhead{HST?} &
\colhead{ID$^a$} &
\colhead{Offset} &
\colhead{Limit$^b$} &
\colhead{Size} &
\colhead{Blend} &
\colhead{Confused} &
\colhead{Notes} \\
\colhead{} &
\colhead{} &
\colhead{} &
\colhead{\arcsec} &
\colhead{\arcsec} &
\colhead{\arcsec} &
\colhead{} &
\colhead{} &
\colhead{}}
\startdata
J123557.1+621725&no&Yang&1.05&0.73&1.7&no &yes&confused by star\\
J123557.7+640831&no&Yang&2.68&1.57&3.9&no&no&\omit\\
J123608.2+621553&yes&HSTz&1.69&0.40&$<0.8$&no&yes&in outer part of galaxy isophotes\\
J123608.3+620852&no&Wang&2.47&0.48&$<1.2$&no&yes&confused by nearby galaxy\\
J123609.1+621104&yes&HSTz&3.58&0.76&$<1.9$&no&no&\omit\\
J123617.4+621442&yes&HSTz&3.92&1.43&$<3.6$&no&no&\omit\\
J123627.7+621158&yes&HST1.6&0.10&0.48&$<1.2$&no&no&\omit\\
J123629.0+621045&yes&HSTz&0.87&0.73&2.2&no&no&aligned with dusty? disk galaxy\\
J123631.2+620957&yes&HSTz&0.56&0.34&$<0.4$&no&no&\omit\\
J123634.6+621421&yes&HST1.6&0.10&0.52&$<1.1$&no &no&\omit\\
\enddata
\tablenotetext{a}{Source of optical/NIR identification used for
  offsets in column 4: Wang: \citet{w10}; Yang: \citet{y14};
  HST1.6,HST1.2, HST814: \citet{k11}; HSTz: \citet{g04}.}
\tablenotetext{b}{maximum offset for ID from equation~\ref{eq1}.}
\end{deluxetable*} 

The parameters of equation~\ref{eq1} have been tuned to limit the
number of random identifications, although it may be possible to
design an even better search strategy. The number of random
identifications was estimated by doing searches offset from each radio
source centroid. The sources which were not identified using  the
equation~\ref{eq1} procedure were then studied further, especially in
the Spitzer bands using the catalog of \citet{y14}, to determine if it
is likely that some of these sources have good identifications as
well.

The important question is whether most of our identifications
  are believable. The answer is complicated by the variable
  sensitivity in the Wang catalog and the potential clustering of the
  NIR sources near the radio sources. In order to mitigate these
  issues and evaluate the identification process globally, we have
  chosen regions near each radio source (as described below), much
  larger than the search areas defined in equation~\ref{eq1} in order
  to evaluate the expected number of false identifications. We believe
  this is a better approach than estimating the probability of the
  correctness of each proposed ID from the NIR source density
  distribution over the entire NIR survey field
  \citep[e.g.][]{ss92,v16}.

\begin{figure}[htb!]
\includegraphics[width=1.03\columnwidth]{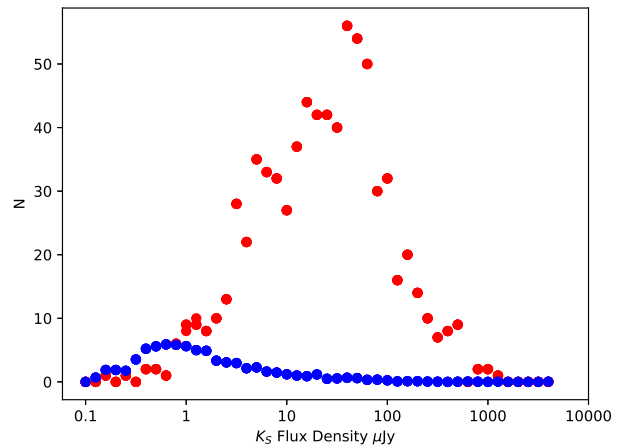}
\caption{Number of  $K_S$ sources  from \citet{w10} versus  $K_S$
  flux density  per logarithmic
    flux density} bin. Red circles are real 
identifications. Blue circles are random identifications.  The
  real identifications are plotted as integers. The random
  identifications have fractional values since the observed values have been scaled
  down from a much larger area searched as described in the text. 
\label{realran}
\end{figure}

In figure~\ref{realran}, we plot the number of sources with a 
  $K_S$ identification within a given  $K_S$ logarithmic flux
density range. Red circles are identifications using the positions and
$radius$ from the values in table~\ref{RC}. Blue circles are the
expected random identifications estimated by searching within
3\arcsec\ of the positions of each source shifted by 10\arcsec\ N,S,E
and W of the true source position.  The random search areas
  where chosen to be near the individual radio sources, but not too
  near, and much larger than areas required by equation~\ref{eq1}.  In
  this way we tried to  minimize the impact of the differences
  in sensitivity while obtaining a useful estimate of the random
  population contaminating the IDs.  Since the random search covers a
  much larger area than the real source search they are scaled down by
  the ratio of the total area searched in the $795\times4$
  3\arcsec\ circles to the sum of the 795 search areas defined by
  equation~\ref{eq1}.

From figure~\ref{realran} one can see that the plot of the  real
  identifications contains a much larger number of objects than 
  expected number of random sources and the real
  distribution is shifted to larger flux densities by $\sim 100$
relative to the randoms. Based on figure~\ref{realran}, we set
  the limiting flux density for IDs in the Wang catalog to $1\mu$Jy,
  although we will examine the sources with non-IDs based on this
  limit further using other data.  $\sim$95\% of the objects in the
survey have a identification with a Ks  source with a flux
density $>1\mu$Jy, near where the random distribution peaks, and
  thus very few of the IDs can be dominated by a random
ID. There are a total of 756 real identifications plotted in
  figure~\ref{realran} with $>1\mu$Jy. The sum of the random IDs 
  $>1\mu$Jy, including the statistical error in the mean, is $27\pm2$
sources or 3.6\% of the real sample. Since these should be distributed
randomly among the real IDs and non-IDs, one would expect $\sim 1.4$ 
(5\% of 27)
of the real IDs with $> 1\mu$Jy to be incorrect, random
identifications.  Furthermore, 95\% of the random identifications $<
1\mu$Jy must be confused with brighter real IDs.

Given that most of the real IDs are brighter than the typical
  random source in any error circle, we pick the brighter NIR candidate
  as the ID in the few cases where more than one object is in the
  error circle. The Wang catalog becomes incomplete below $1\mu$Jy,
so there are more actual weak sources which are confused with a
stronger real source inside our search radius. However, only a few of
the IDs are mischaracterized by identification found in the Wang
catalog.  These faint random sources are mostly blended with the
  real ID, producing small centroid position errors and structure in
  the NIR images of the real ID. The high identification rate 
shown in figure~\ref{realran} compared with the statistical random IDs
confirms that the vast majority of the sources in our radio catalog
are real.

\subsubsection{Nominal non-IDs: further analysis}

 We then considered the
39/795 radio sources that do not have identifications based on the criteria
in the previous section. The properties of these 39 sources are
summarized in table 2.  Did these objects fail to meet
the criteria because
there are no associated objects in the NIR surveys or is there
some other reason ? In order to study this question we have examined
these objects in more detail. All but three have an HST image
  in at least one band.  Eight sources  with HST imaging
need to be excluded due to
various imaging problems documented in table 2.
Six are confused by bright sources on the
best images, including four with HST images, and could have a faint
ID.  Two of the objects with HST imaging appear to be blends, based on
the size and position angle of the radio source and the location of
bright objects in the field.  Two are either confused or the 
  correct identification is slightly further from the radio
  centroid than equation 1 allows.  

Taking into account that
  some of these objects have more than one of the problems listed
  above, this leaves 30 radio sources with HST imaging (and one
without) to be considered. Of these 30, 21 have IDs in at least one of
the HST images, closer than criterion given in equation~\ref{eq1},
mostly much closer. Estimates were made of expected number of random
  sources at any detectable brightness using the same
technique as for the Wang survey for the four different types
of HST images listed in table 2,  \ie\ using 3\arcsec\ radius circles
E, W, N, and S of each source on each type of HST image to estimate
the source density of random IDs for each image type.  
This procedure yielded a total of 0.9 random IDs
expected for all 30 error circles combined. 
This leaves nine radio sources without any ID
based on equation 1 and with HST imaging and one without HST imaging,
 or $\sim 1.3$\% of the  787 radio sources without local problems
  on the images. However, given all the uncertainties discussed and
the expectation of one or two random IDs, it seems most
  likely that less than 2\% of individual radio sources do
not have an ID at the limit of current optical/NIR data.

\begin{figure}[htb!]
\includegraphics[width=1.03\columnwidth]{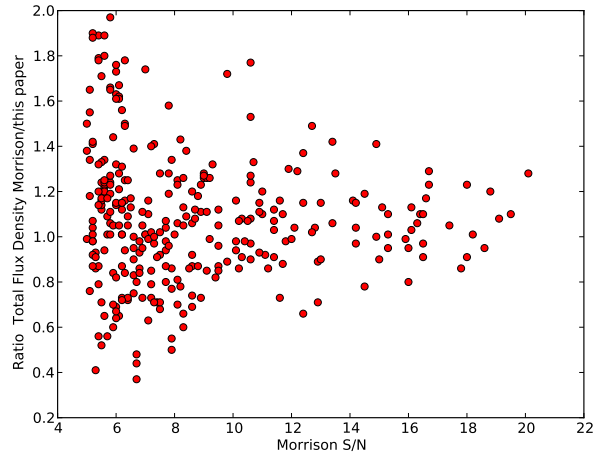}
\caption{The ratio of total flux densities from \citet{m10} versus
  those reported in this paper.  Only sources with $S/N < 20$ in
  \citet{m10} are plotted.
\label{rt}}
\end{figure}

\subsection{Comparison with The Morrison catalog} \label{Mcat}

\citet{m10} published a catalog of GOODS-N using images made from data
from the old VLA,  \ie\ before the JVLA upgrades.  The
  spatial resolutions used to analyze the Morrison catalog were similar
  to the catalog reported in this paper, \ie\ $1.7\times1.6$\arcsec,
  3\arcsec, and 6\arcsec\ in Morrison.
  With the best resolution, the {\it rms} noise
  before the primary beam correction was 3.9$\mu$Jy, as compared with
  2.2$\mu$Jy for our results with the JVLA. Inside the 9\arcmin\ radius circle
  reported for our new results, there are 484 sources in the Morrison
  catalog, compared with our 795.
  In addition, the Morrison data were forced by
the parameters of the old VLA correlator to use a much narrower total
bandwidth, channel widths of 3.125 MHz and an averaging time of 3
seconds. The individual channels also were almost the same
  width but not exactly the same shape. These parameters created
smearing of the image which increases as one goes away from the
pointing/phase center. These effects were approximately taken into
account in the Gaussian fitting process but made measurement of the
sizes of the sources and thus the total flux densities more difficult.
Furthermore, the total bandwidth used in the imaging was
only 44 MHz. Besides reducing the sensitivity, the narrow bandwidth
produces poorer uv-coverage and thus higher sidelobes for the dirty
beam than the new JVLA data. 

 The Morrison survey did include A, B, C, and D configuration
  data. However, no very large sources were found. The VLA
  A-configuration has uv-coverage adequate for minor axis scales up to
  36\arcsec\ , so based on the sources found in the Morrison survey,
  the B, C, and D configurations were {\bf not} included in this survey.

The median position difference in the sense JVLA-Morrison is about
0.2\arcsec\ in RA and 0.0\arcsec\ in Declination. Since the Wang K-band survey
was referenced to the Morrison frame of reference, we have corrected
the ID position delta for this difference.  It is not clear where the
RA difference arises.

\begin{figure}[htb!]
\includegraphics[width=1.03\columnwidth]{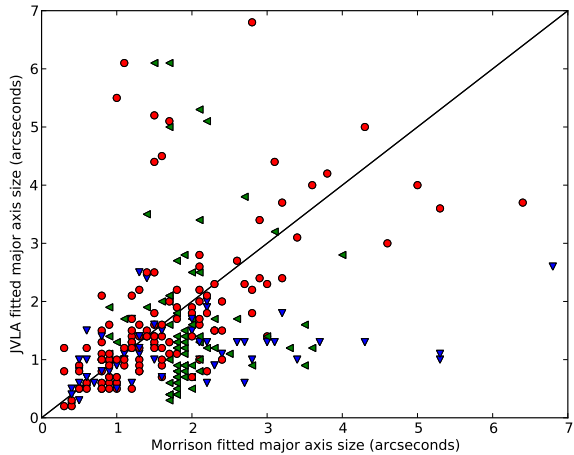}
\caption{Gaussian major axis sizes from \citet{m10} versus those
  reported in this paper.  Only sources with both major axis sizes
  $<7.0$\arcsec\ and a non-upper limit size in one of the two catalogs
  are plotted. Sources for which both catalogs find the source is
  resolved are plotted as red dots. Those sources with upper limits in
  one catalog are plotted as a blue (Morrison catalog) or green
  triangles (new VLA catalog) with one point of the triangle in the
  direction of the limit.
\label{smn}}
\end{figure}

In figure~\ref{rt}, we show the ratio of the \citet{m10} total flux
densities to the total flux densities for each source in common
between the two surveys with Morrison $S/N <20$. The scatter in the
ratio clearly increases at low Morrison S/N.   However the median ratio
between the two catalogs is 1.06, which is very consistent with the
slight difference in the effective frequency of the observations
(1400 MHz for \citet{m10} and 1525 MHz for this survey reported
here),  given the internal errors, the counting statistics and a
  mean spectral index in the expected range of $-0.7$ to $-0.8$.
Assuming the higher S/N data reported here are correct, it appears
that in some cases at low S/N \citet{m10} finds sources that are
spuriously resolved and in other cases misses extended structure. This
is just what one might expect at the faint end or ``bottom'' of a
catalog,  \ie\ near the $5\sigma$ limit and is consistent with
the simulations described earlier.

In figure~\ref{smn} the fitted major axis sizes for sources in both
the catalogs are plotted.  Only sources found to be resolved in one of
the two catalogs are plotted. The plot shows a good underlying
correlation between the sizes found,  especially if one focuses on the
points with measured values in both catalogs, \ie\ the red points.  As
expected the new results find a few sources in the upper left of the
plot which are much larger in the new catalog,  presumably due to
the lower noise and the more extensive search done for resolved
sources, \ie\ using six resolutions as described in
  sec.~\ref{cc}).  
However, there also is a much larger set of sources
significantly below the equal size line. Some of these are upper
limits in the Morrison catalog but a significant population is present
for which \citet{m10} found a larger size.  A lot of these differences
are likely due to the tuning of EFACTOR as discussed above, which was
not done for the \citet{m10} results. The other imaging artifacts, due
to the old VLA correlator and the longer averaging times required for
the earlier epoch data, must contribute to the differences.

\begin{figure*}[htb!]
\epsscale{0.98}
\plottwo{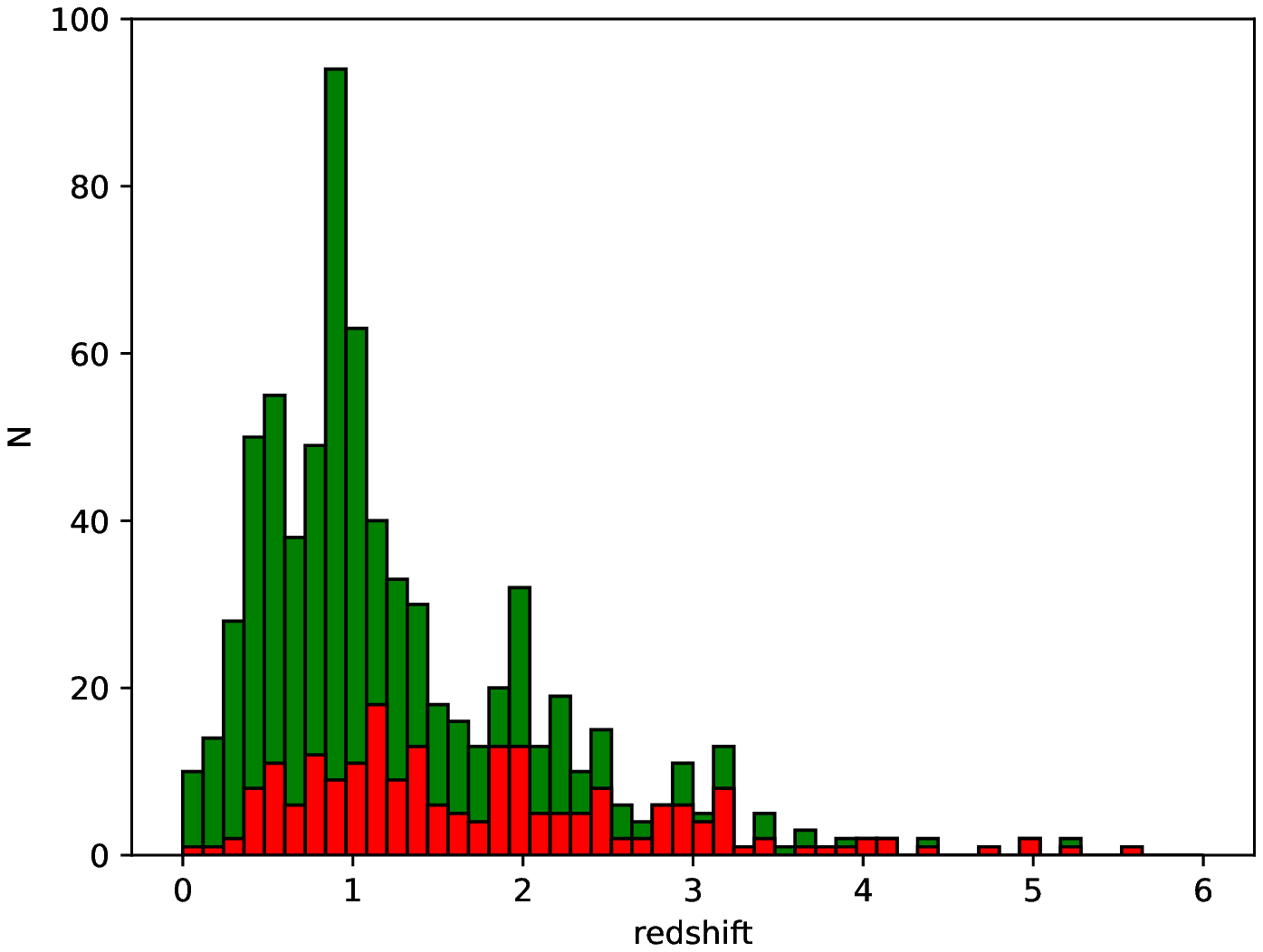}{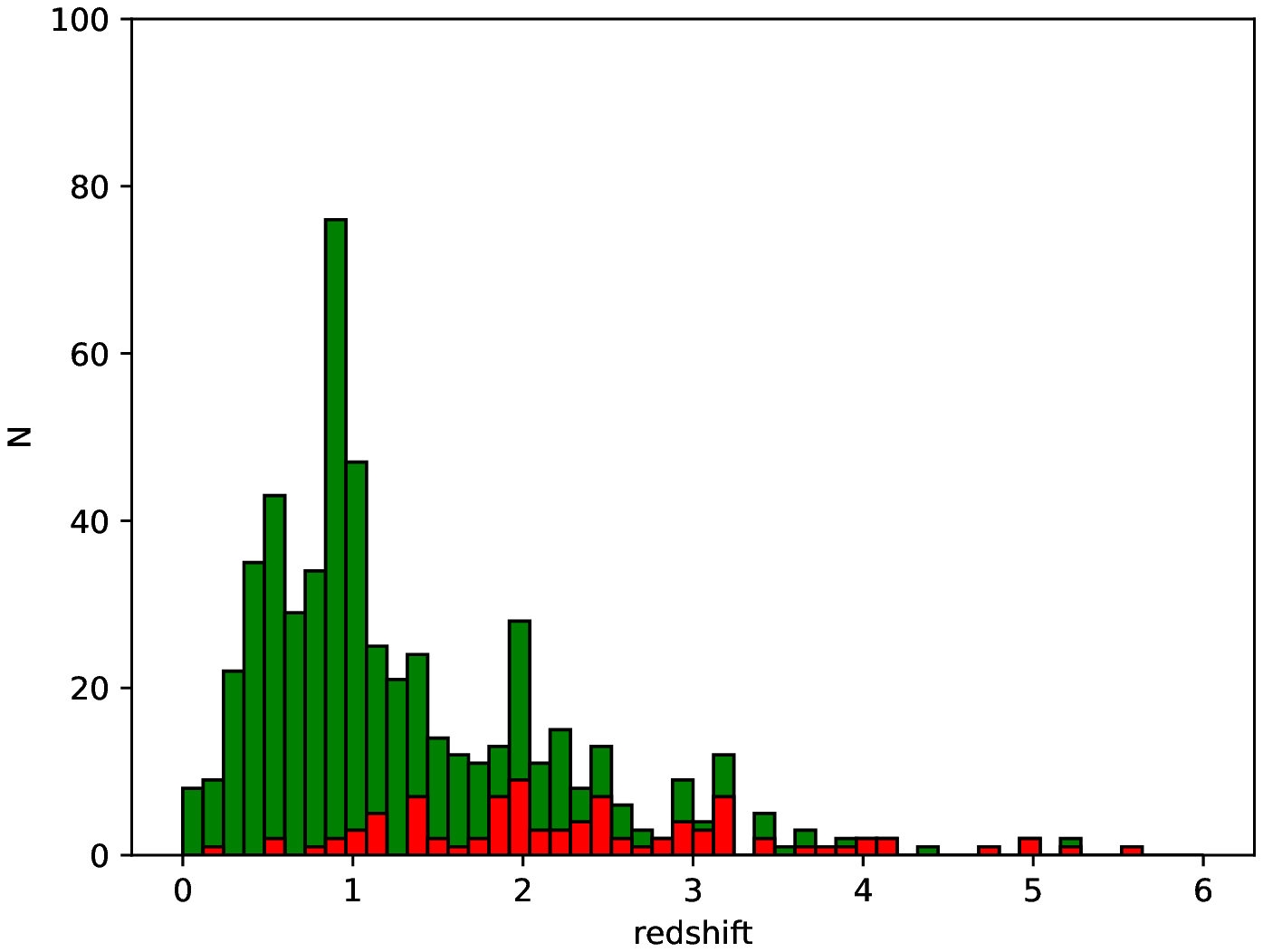}
\caption{Redshift histograms for full radio survey (left plot)
  and  radio survey with z-band HST imaging (right plot). Green indicates
  spectroscopic and red indicates photometric redshifts.
 \label{zhist}} 
\end{figure*}

\begin{figure*}[htb!]
\epsscale{0.98}
\plottwo{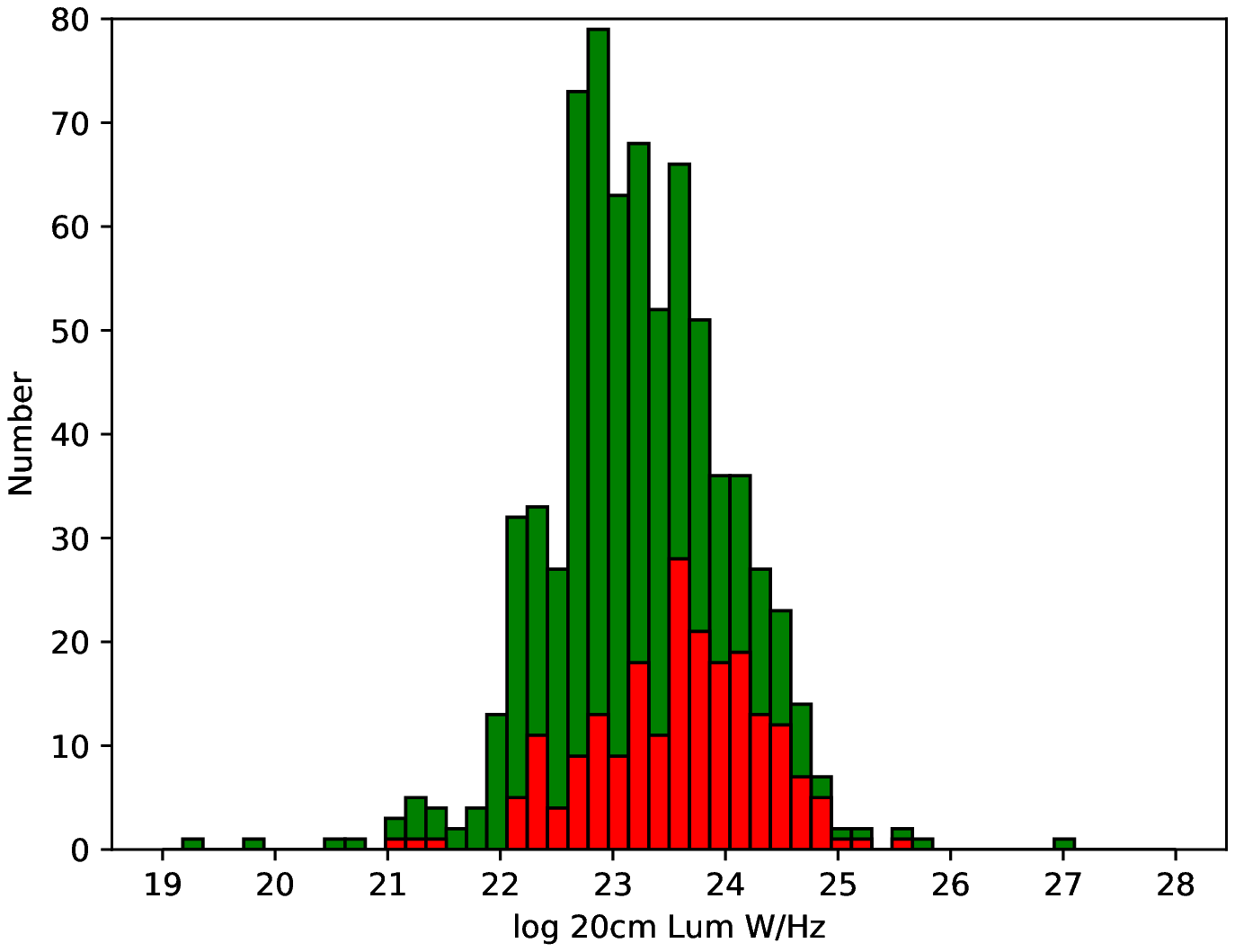}{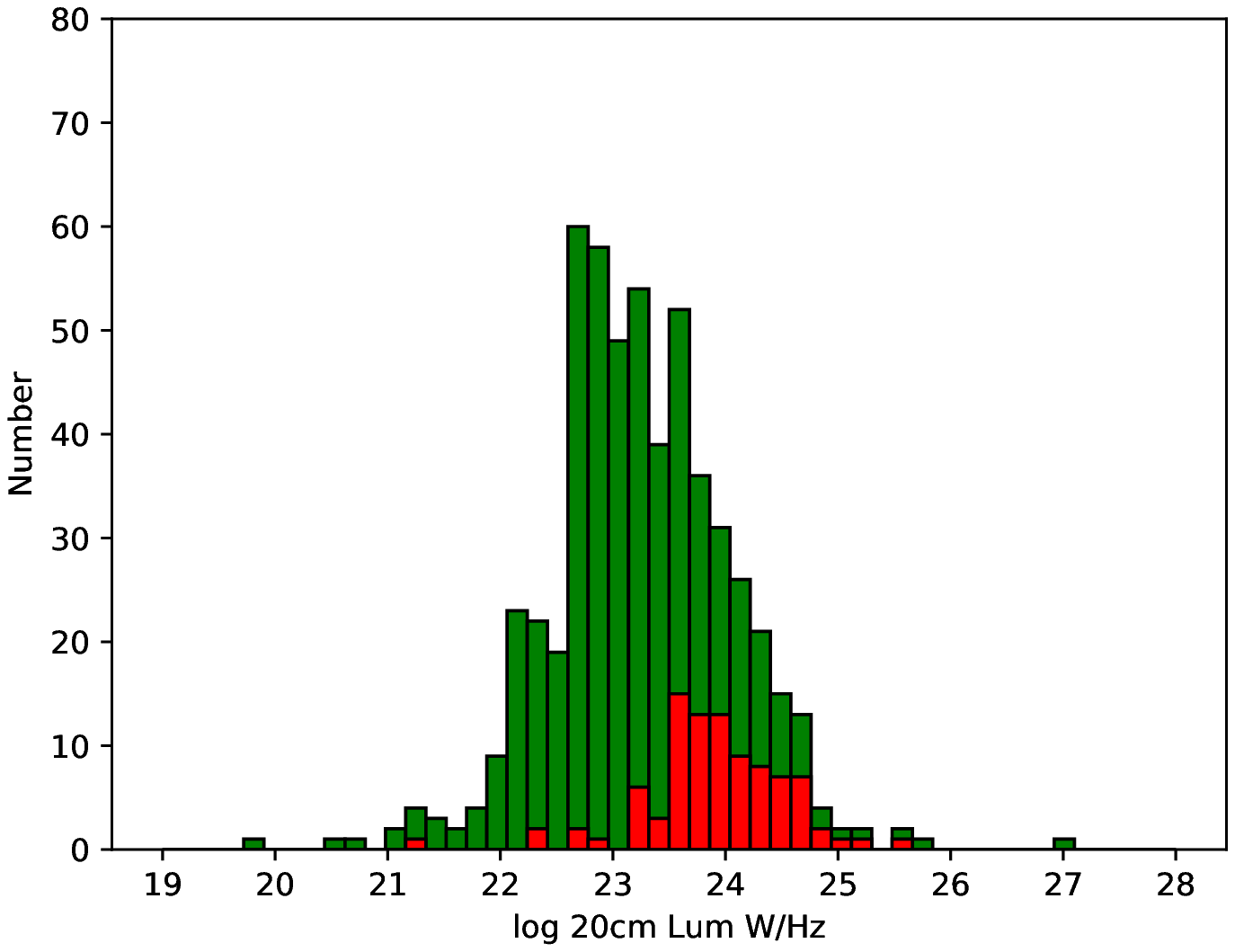}
\caption{Radio luminosity histograms for full radio survey (left
    plot) and radio survey with z-band HST imaging (right
    plot). Green indicates spectroscopic and
  red indicates photometric redshifts. 
 \label{lumhist}}
\end{figure*}

At the ``bottom'' of a catalog, \ie\ as one approaches $5\sigma$ detection
limit, the ability to measure any information in a given image beyond the
peak becomes limited. Unlike \citet{m10} in the new catalog reported
here very few sources have size estimates near this limit as a result
of our tuning of the fitting parameters (see fig~\ref{size} and
section~\ref{RS} below).
 
\subsection{Redshifts, Luminosities}

One of the principle reasons to study GOODS-N is the large number of
redshifts available for this field. Especially in region of the
original GOODS-N HST survey, the number of spectroscopic redshifts is
very large. In table~\ref{RC} the redshifts used and their origin are
listed.  In figure~\ref{zhist}, we plot the histogram of measured
redshifts for the full survey and for the subset with z-band HST
imaging. The vast majority have spectroscopic redshifts, especially
the HST z-band subset. Both redshift distributions peak at $z\sim1$
with a long tail to higher redshifts.

 The completeness
  level for spectroscopic redshifts, especially for the sample with
  HST z-band imaging, is very good. In the full survey, there are
  520 sources with spectroscopic redshifts, 210 with photometric
  redshifts and 65 either without redshifts or with no ID. Of those
  sources with HSTz images, there are 465 with spectroscopic
  redshifts, 92 with photometric z's and 26 with no redshifts
  including those with no ID.  In both samples, a large fraction of
  the $z>2$ redshifts are still based on photometric redshifts.

  The corresponding 20cm radio luminosities are plotted in
  figure~\ref{lumhist}, assuming $H_0=69.6$ $\Omega_M=0.286$,
  $\Omega_{vac}=0.714$.  The photometric redshifts contribute a
  significant fraction of the HST sample at $z\gtw 10^{24}$ W
  Hz$^{-1}$. Generally, comparison of the redshifts for radio
    sources with both photometric and spectroscopic redshifts show
    that the two estimates agree very well. Also see \citet{y14} from
    which almost all of the photometric redshifts where taken for a
    general comparison. However, for the $z >2$ objects, we still
  have few spectroscopic results. This limits our ability to use
  spectral lines to help us distinguish between AGN and star-formation
  as the origin of the radio emission and makes us more dependent on
  the photometric results for z. The corresponding radio luminosities
  at which we start to rely on photometric redshifts are also where we
  might expect a large AGN contribution or perhaps evolution-driven
  large star-formation rates.  So for the higher redshift part of the
sample, interpretation becomes harder.  For local galaxies, say
$z<0.2$, 20cm radio emission is dominated by star-formation for
$L_{20}\ltw 10^{23}$ W Hz$^{-1}$ and by AGN for higher radio
luminosities \citep{c02}.  For star-formation $10^{23}$ W Hz$^{-1}$
corresponds roughly to a star-formation rate of 60 M$\odot$/yr
\citep{y01,m11}, slightly below lower limit for ULIRGs
( $L_{FIR}=10^{12} L_\odot \sim100$
M$\odot$/yr; $log(L_{20})\sim23.25$).  A local peak in the histograms
shown in figure~\ref{lumhist} can be seen, corresponding to 
$L_{20}\ltw 10^{23}$ W Hz$^{-1}$ and star-formation rates mostly
between 3 and 20 M$\odot$/yr, as is found for low redshift
galaxies.  Based on this calibration, the majority of the
objects in figure~\ref{lumhist} must correspond to AGN, or LIRGs
 ($\sim 10-100$ M$\odot$/yr) and
ULIRGs with SFR$\gg20$ M$\odot$/yr.

\subsection{Radio Sizes} \label{RS}

\begin{figure}[htb!]
\includegraphics[width=1.03\columnwidth]{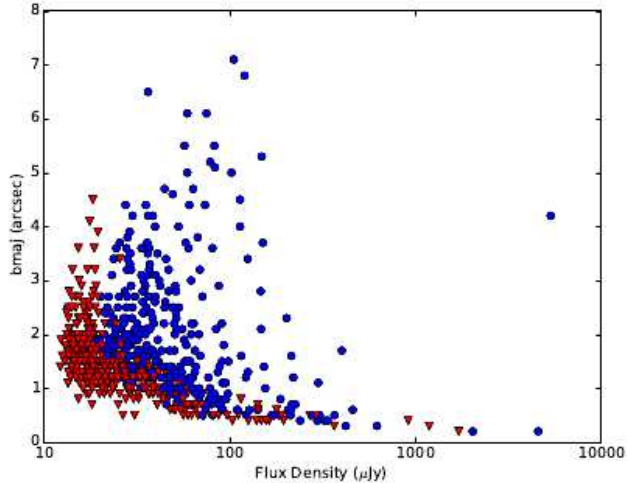}
\caption{ Radio Major Axis Size  (bmaj) versus 
    Radio Total Flux Density. Red downward arrows are sources with major axis
upper limits. Blue circles are sources with significant measured major axis sizes.
\label{size}}
\end{figure}

In table 3 we give  the list of the deconvolved, fitted Gaussian
sizes for each source. In figure~\ref{size} we show a plot of these
sizes versus total flux density. 346/795 sources, almost half of
  the total, are found to be resolved.  274 of these have
  deconvolved resolved sizes with major axes $>1$\arcsec\ and 72 have
  resolved major axes $\le1$\arcsec. 174 have either upper limits
  $\le1$\arcsec\ (72) or measured resolved sizes, not upper
  limits, $< 1$\arcsec (102).
  Since only a minority of the sample have high enough S/N to produce
  a resolved fit $<1$\arcsec, we do not try to study these small
  sources in statistical detail. The remaining 347 sources have upper
  limits $>1$\arcsec, we also cannot study the sizes in detail at
  least at the 1\arcsec-level. Thus in the rest of this paper
  we focus mostly on the resolved sources with major axes $>1$\arcsec\ .
  
The survey is also necessarily incomplete in total flux density due to
the small size of the $1.6$\arcsec\ clean beam most often used to set
the size limits for weak source, especially at the bottom of the
catalog where the S/N is low. Almost all the sources with total flux
densities $< 20\mu$Jy are not formally resolved (see
figure~\ref{size}), suggesting that some resolved sources in this
lower flux density range could be missing and some of the listed
sources likely have larger structure. The images were searched
for sources at lower resolution but the convolved images have higher
noise levels than the 1.6\arcsec\ image. At higher flux densities, the
plot shows sources with extended sizes up to $\sim 7$\arcsec, although
the median size below $100\mu$Jy is $< 1.5$\arcsec. The median source
size for total flux densities between 10 and 40$\mu$Jy is an upper
limit, $<1.7$\arcsec\ . In the 40 to 160 $\mu$Jy range, the median
size is 1.3\arcsec\ .

\subsection{Comparison of Radio and HST Sizes and Orientations}

\subsubsection{Radio versus HST Sizes}

\begin{figure*}
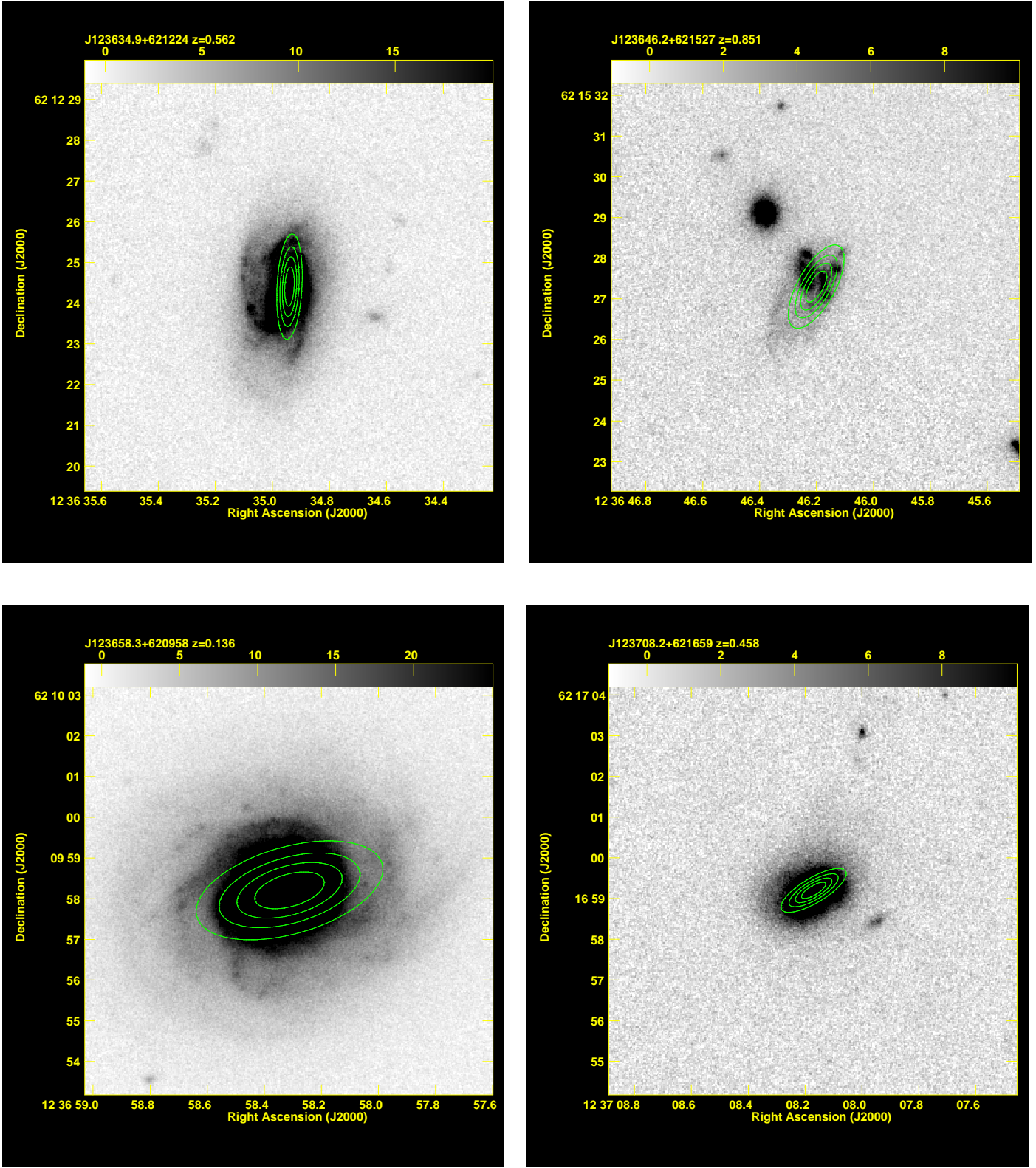


\includegraphics[width=1.03\columnwidth,angle=0, trim=0cm 0cm 0.0cm 0.0cm,clip=True]
{J123634TVC1GNOD.ps}
\hspace{0.035\columnwidth}
\vspace{0.02\columnwidth}
\includegraphics[width=1.03\columnwidth,angle=-0,trim=0cm 0cm 0.0cm 0.0cm,clip=True]
{J123646TVC1GNOD.ps}
\vspace{0.02\columnwidth}
\\
\includegraphics[width=1.03\columnwidth,angle=0,trim=0cm 0cm 0.0cm 0.0cm,clip=True]
{J123658TVC1GNOD3.ps}
\hspace{0.03\columnwidth}
\includegraphics[width=1.03\columnwidth,angle=0,trim=0cm 0cm 0.0cm 0.0cm,clip=True]
{J123708TVC1GNOD3.ps}
\caption{HSTz images with radio models overlaid from table~\ref{RC}
  for Individual
  Galaxies identifications. Fractional contours of $(0.2,0.4,0.6,0.8)\times$
  the model peak are plotted.
\label{fourway_G}}
\end{figure*}

\begin{figure*}
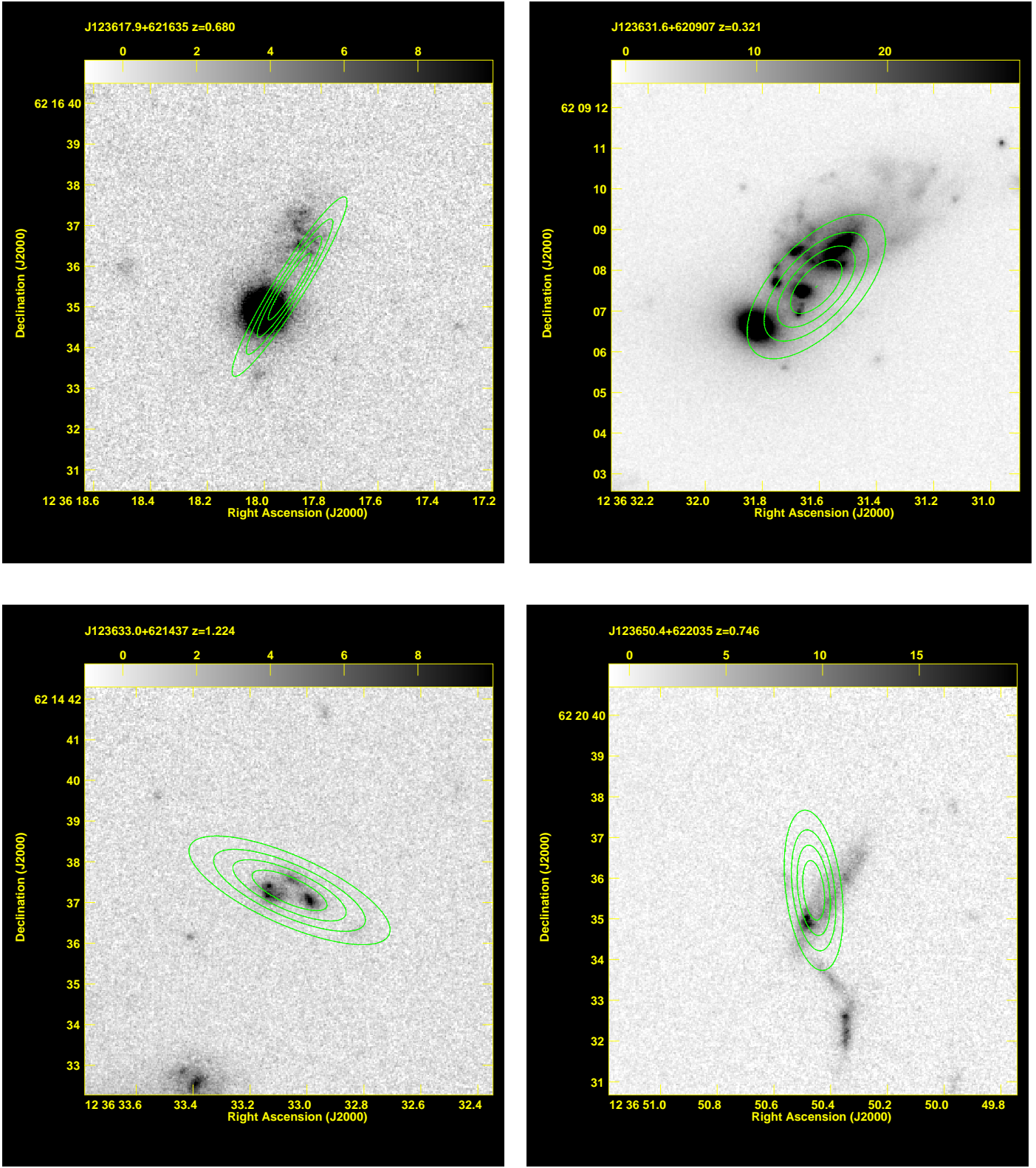


\includegraphics[width=1.03\columnwidth,angle=0, trim=0cm 0cm 0.0cm 0.0cm,clip=True]
{J123617TVCB.ps}
\hspace{0.035\columnwidth}
\vspace{0.02\columnwidth}
\includegraphics[width=1.03\columnwidth,angle=-0,trim=0cm 0cm 0.0cm 0.0cm,clip=True]
{J123631TVCB.ps}
\vspace{0.02\columnwidth}
\\
\includegraphics[width=1.03\columnwidth,angle=0,trim=0cm 0cm 0.0cm 0.0cm,clip=True]
{J123633TVCB.ps}
\hspace{0.03\columnwidth}
\includegraphics[width=1.03\columnwidth,angle=0,trim=0cm 0cm 0.0cm 0.0cm,clip=True]
{J123650TVCB.ps}
\caption{HSTz images with radio models overlayed as contours for Binary 
    or Multiple Galaxy identifications (ID ``B'' in table 3). Fractional
  contours of $(0.2,0.4,0.6,0.8)\times$ the model peak are plotted.
\label{fourway_B}}
\end{figure*}

For the resolved radio sources it is important to compare their
properties with the HST optical imaging  \citep{g04,k11}.  
  In figures~\ref{fourway_G} and~\ref{fourway_B}, we display examples
  of the fitted radio models from table~\ref{RC} overlaid on the HSTz
  images. The models, of course, have errors and are an approximation
  to the true structure so are not expected the line up
  perfectly. However, one can see the similarity to the optical
  structure.

  For the binary systems, the sizes tend to be small,
  supposedly because the larger binaries would likely be measured as
  two separate identifications. For the multiple systems the
  measured size is the distance between the centers of the outer
  components of the system. The models are an
  approximation to the actual, often much more complicated brightness
  distribution. In figure~\ref{fourway_B}, J123633.0+621437 is an
  example of a fit that suggests the brightness distribution extends
  beyond the three galaxies. On the other hand, the model for
  J123650.4+622035 is centered near the galaxy near the field center
  but is not aligned with that galaxy. Instead the alignment fits with
  the PA of more southern components of the multiple system. We have
  interpreted the extended source as due to the multiple system with
  some unknown more complex structure, instead of being an extended
  misaligned source associated with a single galaxy.  

For sources
  with HSTz data sizes $> 1$\arcsec\ (169 sources), we have
measured the largest isophotal size of the galaxy and multiple/binary
galaxy IDs on the HSTz images, \ie\ the isophote at which the galaxy
brightness blends into the noise. This ``lowest'' isophote was used
  for consistency and because the lowest isophote gives the largest
  lever arm for measuring the orientation of the galaxy. In practice
  the vast majority of the galaxies had very similar position angles
  for the inner and outer structure.

\begin{figure}[htb!]
\includegraphics[width=1.03\columnwidth]{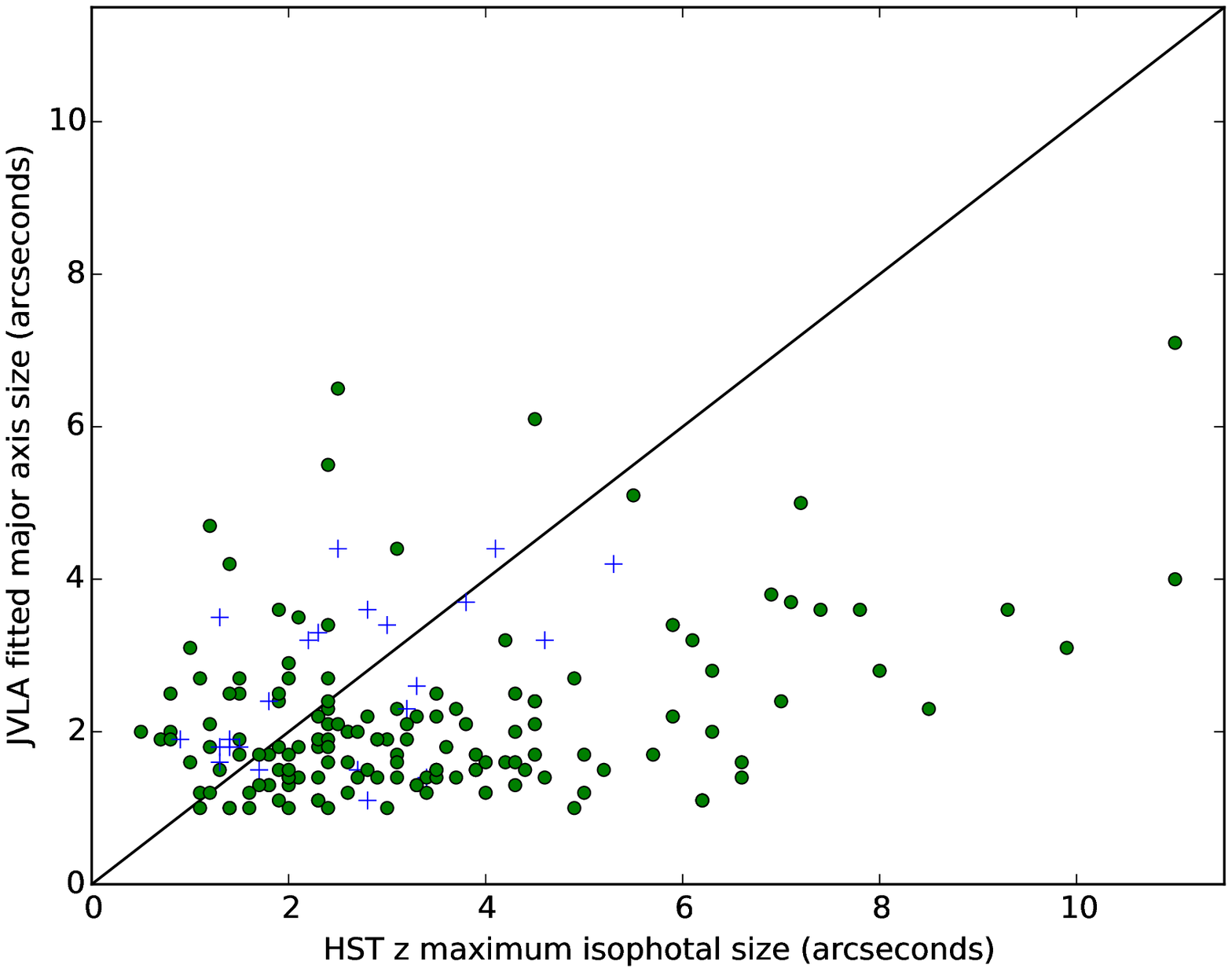}
\caption{VLA deconvolved Gaussian major axis (FWHM) versus maximum isophotal
  HST z-band major axis for radio sources with sizes $>1\arcsec$ identified
  with individual galaxies (green circles) and binary systems (blue pluses).
\label{ros}}
\end{figure}

   In table~\ref{Ang} we list the measured radio and HSTz optical
  sizes and position angles for the 169 radio sources with deconvolved
  major axis sizes $>1$\arcsec.  In figure~\ref{ros} we plot the
radio versus optical sizes for those 169 sources.  In most cases the
individual optical galaxies measured this way are larger than the
radio sources by about $1.5\times$ in the median. For multiple/binary
optical systems are slightly smaller than the radio major axes, $\sim
0.8\times$ in the median. Note that we are measuring very different
quantities for the size, so we should not expect perfect
agreement. The radio Gaussian size is a proxy for the 2nd moment. The
optical isophotal size is a function of redshift due to the
cosmological $(1+z)^4$ isophotal dimming and the redshifting of the
z-band below the 4000\AA\ break for $z \gtw 1$.  Most of the galaxies
appear to be disks but not all, so the significance of the size
measured from outer isophote will be different for the early-type
systems.

\begin{deluxetable}{llrrlrll}[htb!]
\tabletypesize{\scriptsize}
\tablecolumns{8}
\tablewidth{0pt}
\tablenum{3}
\tablecaption{Radio and HSTz optical position angles 
 for sources with radio major axes $>1$\arcsec}$^a$\label{Ang} 
\pagestyle{empty}
\tablehead{
\colhead{Name} & 
\colhead{ID} &
\colhead{R PA} &
\colhead{Err} &
\colhead{Res} &
\colhead{O PA}&
\colhead{R Si}&
\colhead{O Si}\\
\colhead{(1)} &
\colhead{(2)} &
\colhead{(3)} &
\colhead{(4)} &
\colhead{(5)} &
\colhead{(6)} &
\colhead{(7)} &
\colhead{(8)}}
\startdata
J123541.4+621217&G&133&16&1.6&125&3.2&4.2\\ 
J123546.7+621048&G&78&25&1.6&98&1.6&2.4\\ 
J123553.1+621073&G&146&15&1.6&6&1.4&2.1\\ 
J123553.1+620954&B&161&41&1.6&24&1.4&3.4\\ 
J123554.0+621043&G&150&17&1.6&146&1.1&6.2\\ 
J123558.1+621355&G&110&31&6.0&119&6.1&4.5\\ 
J123559.7+621549&G&44&8&1.6&39&1.6&6.6\\ 
J123600.4+621053&G&25&39&1.6&51&2.5&0.8\\ 
J123601.1+621058&G&13&15&1.6&172&1.9&3.0\\ 
J123605.4+621031&G&39&11&1.6&30&2.7&2.4\\ 
\enddata
\tablenotetext{a}{Col 1: Name, Col 2: ID type (G: Individual Galaxy, B:
  Binary or Multiple Galaxy), Col 3: Radio position
  Angle, Col 4: Error radio position angle from JMFIT, Col 5:
  Resolution of Image used for Radio Position angle, Col 6: Optical
  position angle from HSTz image, Col 7: Radio major axis size, Col 8:
  Optical major axis size}
\end{deluxetable}

A small number of galaxies have smaller optical sizes than the radio
sources. About two thirds of these have $z>1$ and are small, $<1.5$\arcsec,
perhaps due to the redshift dependent effects discussed above.
 At $z>1$ the radio luminosities are also larger, $> 10^{23}$ W
Hz$^{-1}$, thus some may be AGN. One in particular, J123644.4+621133,
is $43$\arcsec\ in radio size and is clearly an FRI. 

\begin{figure*}[htb!]
\epsscale{0.98}
\plottwo{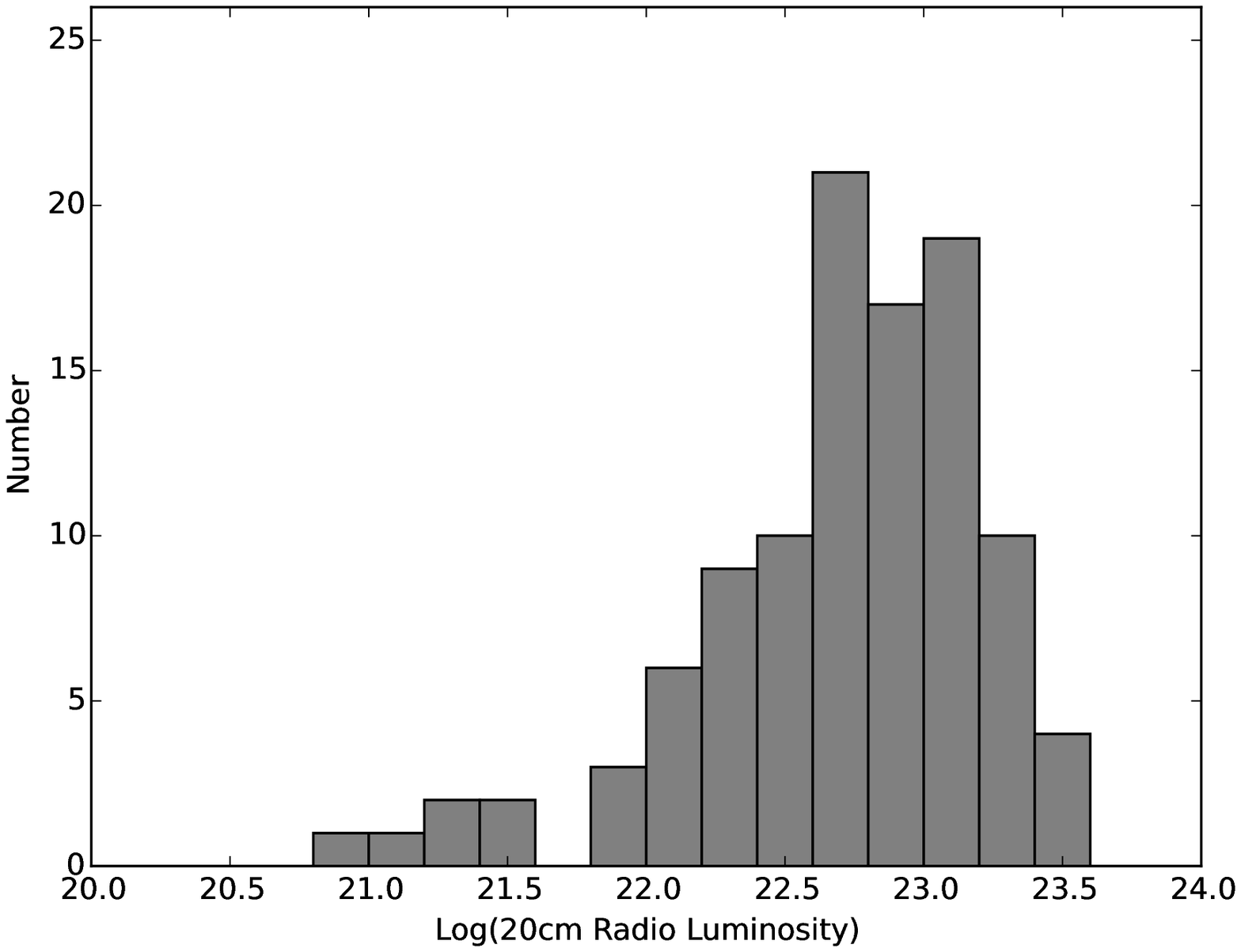}{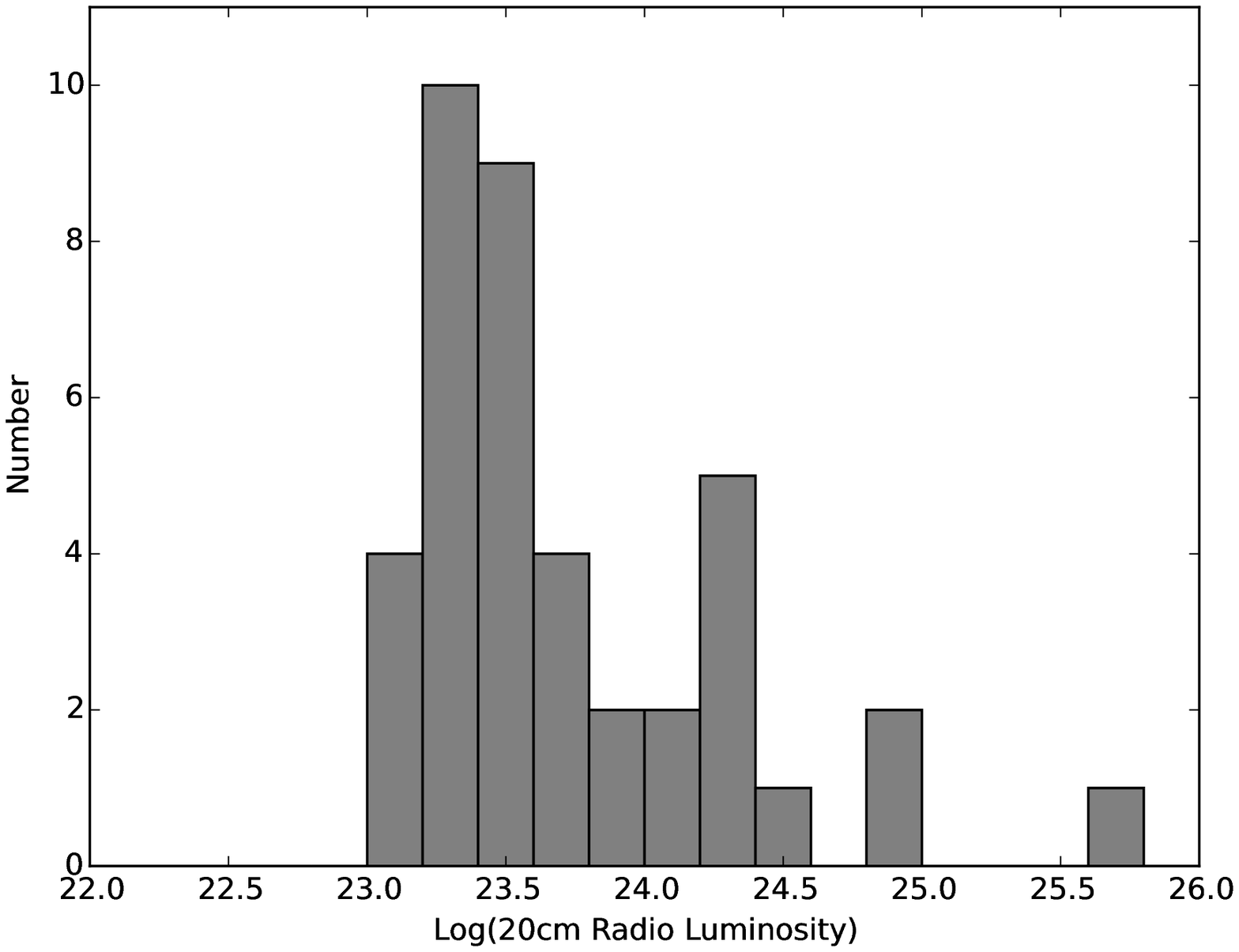}
\caption{  Radio Luminosities for sources with HST z imaging and radio
  sizes $>1\arcsec$: left) $z<1$ (left plot), $z>1$ (right
    plot). Note the split
  between the two samples very near  $L_{20}=10^{23}$ W
  Hz$^{-1}$. Locally $10^{23}$ W Hz$^{-1}$ is the dividing point
  between dominance by star-forming galaxies and dominance by AGN.
 \label{gzlum}} 
\end{figure*}

\begin{figure*}[htb!]
\epsscale{0.98}
\plottwo{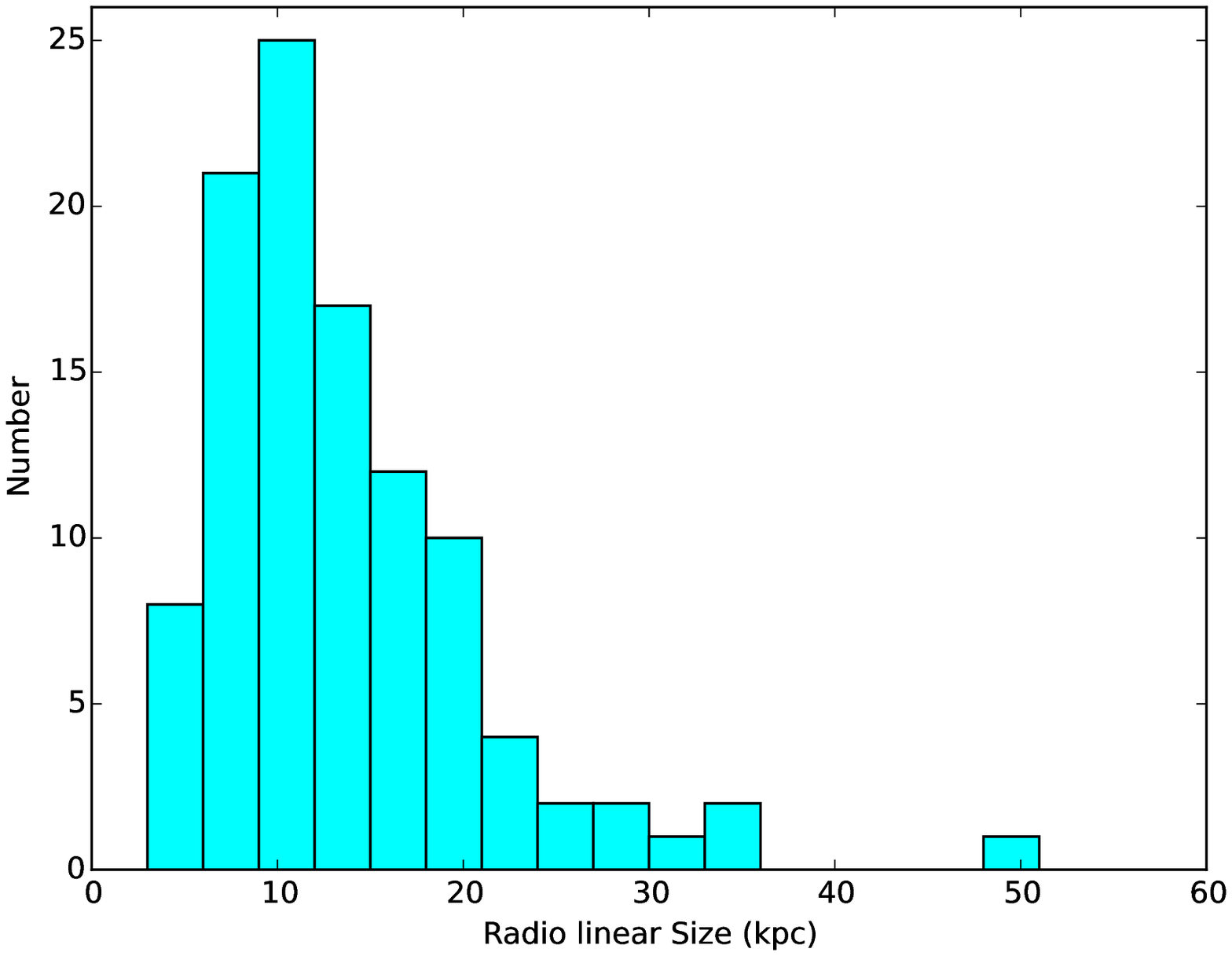}{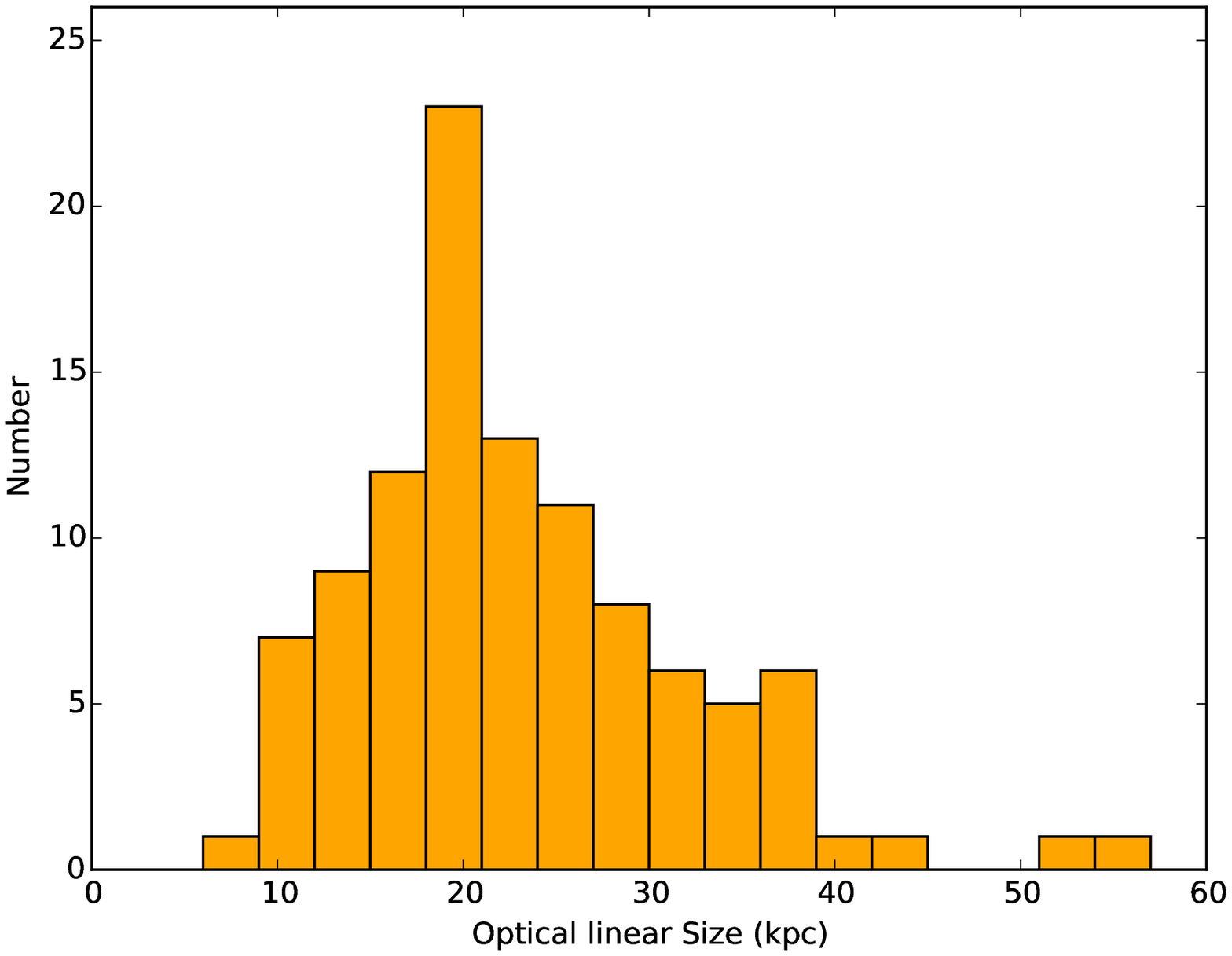}
\caption{ Radio Galaxies with HST z imaging, sizes $>1\arcsec$ and $z<1$:
  radio Gaussian major axes (left plot),  optical isophotal major
  axes (right plot). The optical sizes show that these galaxies are much larger than
    local ULIRGs, suggesting that that the radio size selection picks out such
  monolithic systems. The radio sources, while larger than the nuclear
  region, are systematically smaller than the optical isophotal size,
  showing they are contained inside their associated galaxies.
 \label{grosizelt1}} 
\end{figure*}

\begin{figure*}[htb!]
\epsscale{0.98}
\plottwo{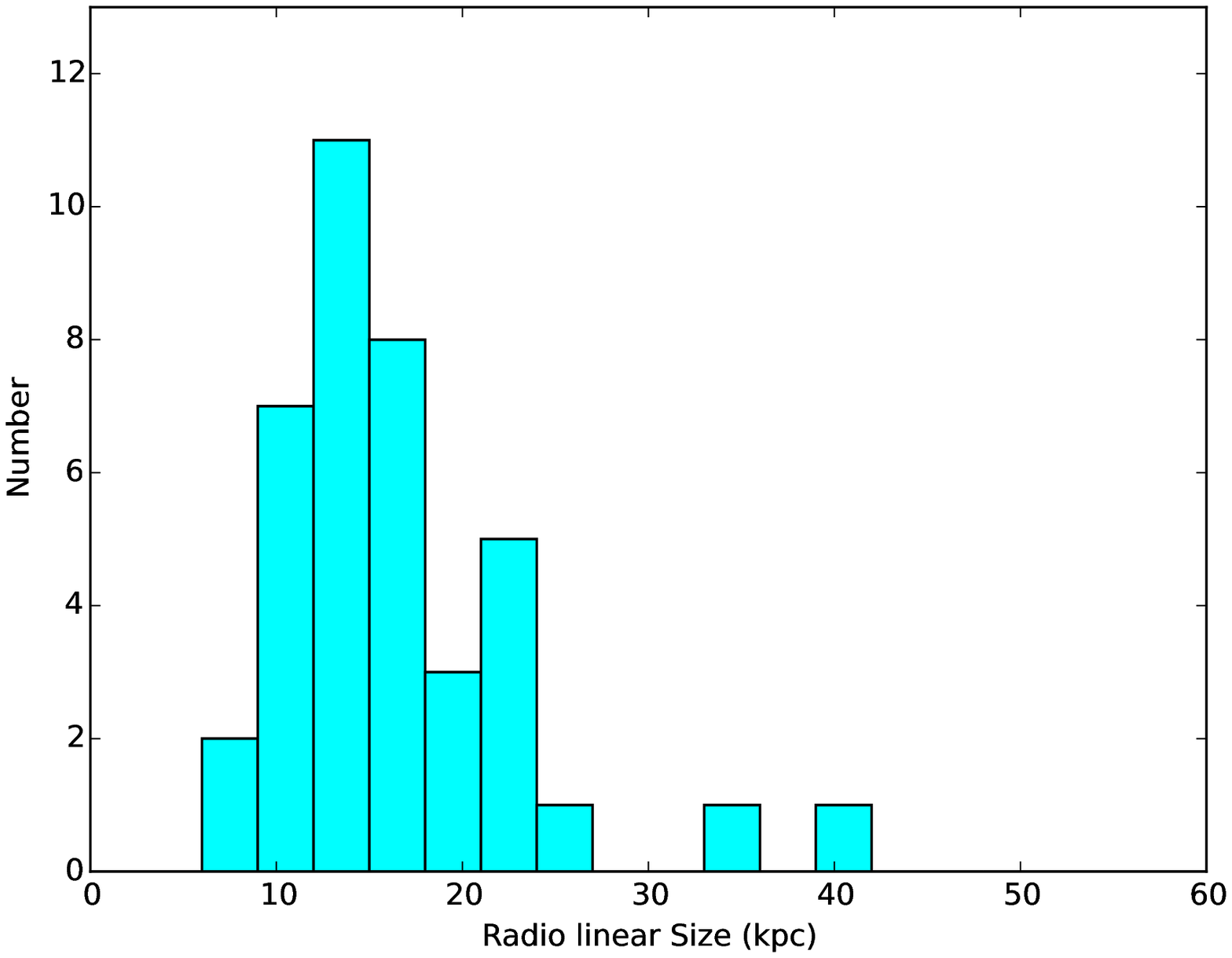}{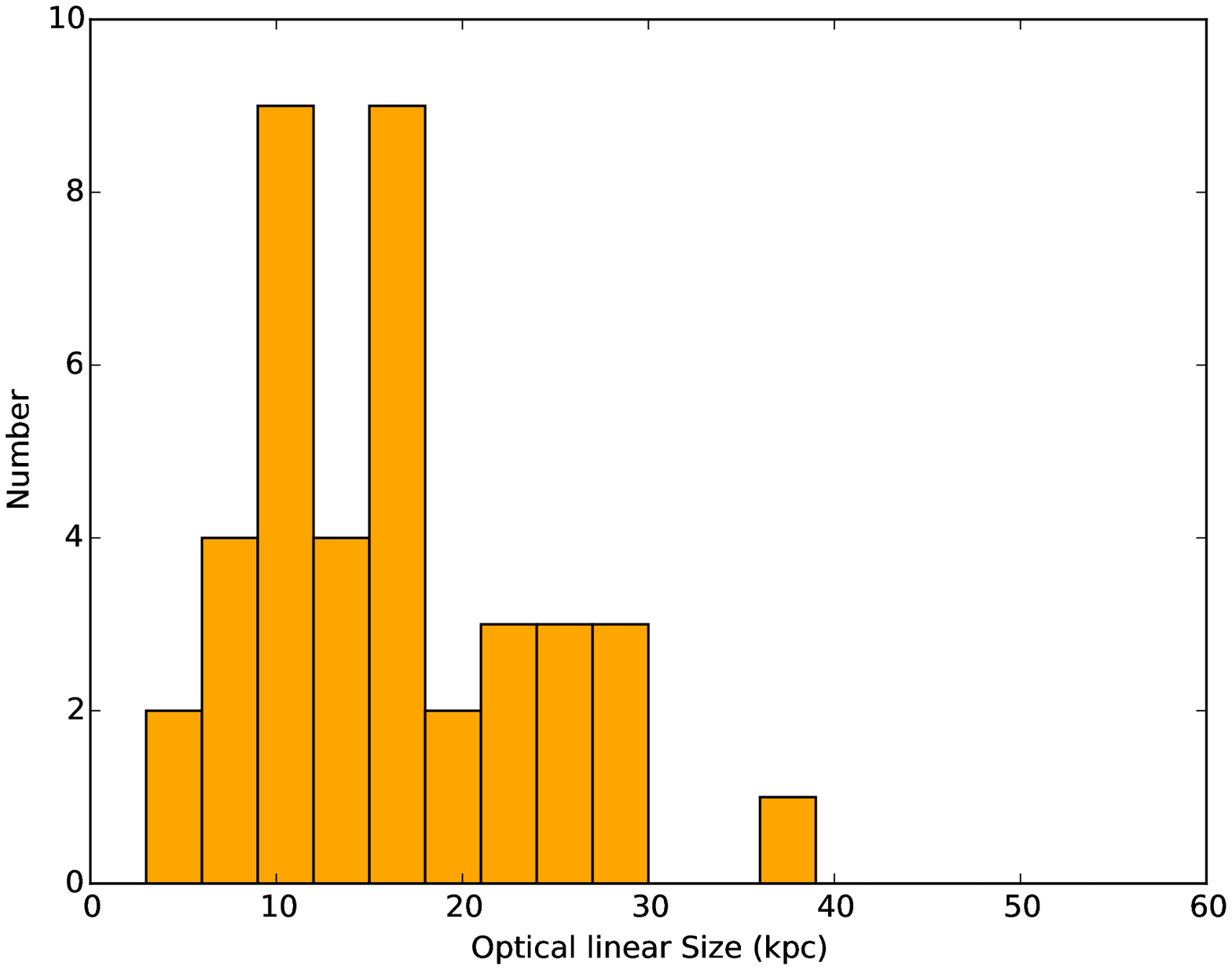}
\caption{  Radio Galaxies with HST z imaging,  sizes $>1\arcsec$ and $z>1$:
 radio Gaussian major axes (left plot) and optical isophotal major
  axes (right plot). For $z>1$ unlike $z<1$, the radio and optical sizes are similar, perhaps
  due to the cosmological effects on the optical isophotes. 
 \label{grosizegt1}} 
\end{figure*}

In figure~\ref{gzlum} histograms of the radio luminosity are plotted
for galaxies with sizes $> 1$\arcsec\ splitting the population into
$z<1$ and $z>1$. This divides the sample by luminosity very close to
$log(L_{20})=23$, which locally is the dividing line between the vast
majority of the radio sources being star-forming for lower luminosities
and being AGN for higher luminosities \citep[e.g.][]{c02}.

For sources identified with individual galaxies with sizes $>
1$\arcsec, 107/147 have $z < 1$, where $z=1.01$ is the median redshift
for the radio sources so far measured (the vast majority).  Thus
 these resolved individual galaxies tend to be at the lower
redshifts, corresponding to lower radio luminosities. The median
redshift for the $z<1$ sample is 0.56 and the median
$log(L_{20})=22.77$. Assuming star-formation is the origin of the
radio emission, most of the objects are in the LIRG class
and about
10\% extend into the ULIRG class. Forty individual galaxies are found
with $> 1$\arcsec\ and $z>1$.  The redshifts are mostly $<1.5$ with a
tail extending up to almost $z=4$. Assuming the local star-formation
relation, radio luminosities are all in the ULIRG class.

In figure~\ref{grosizelt1} we show histograms of the radio and optical
sizes from table 4. For $z<1$ the radio sizes have a median of 11.6
kpc, while the median optical size is 21.5 kpc. Thus the radio
emitting region extends well outside the galaxy nucleus but is
typically less than than half our measured optical isophotal
size. These galaxies, selected by picking radio sources
  $>1$\arcsec\ , are large, grand-design spirals or disk galaxies with
  isophotal diameters up to 60kpc. They are unlike the local population
  of ULIRGs, which are much smaller, disturbed systems dominated by nuclear
  starbursts and the associated compact radio emission
  \citep[e.g.][]{c91,r11,bm15}.  While this is a subset of the $\sim
400$ sources with $z<1$, it is a large subset and shows that a
significant fraction of sources have extended emitting regions outside
the nucleus. Furthermore, typically the unresolved sources have upper
limits to the radio size, $\sim 10$ kpc. Thus it is likely that
  more of the galaxies have extended emitting regions outside the
  nucleus than the subset we have identified.  The vast majority of
the identifications appear to be disk galaxies, thus it appears that
 many of the disks are emitting and that the star-formation is
not limited to a nuclear star-burst for a significant fraction of the
$z<1$ galaxies in the sample.

For 39 radio galaxies with $> 1$\arcsec\ and $z>1$, we display their
radio and optical size in figure~\ref{grosizegt1}. One object,
J123644.4+621133, is not plotted since it is too big to fit on the
radio size plot and is clearly an AGN based on its radio morphology.
The radio and optical sizes are similar with a median near 15 kpc. Of
course, the optical sizes are affected by the isophotal dimming due to
both cosmological effects and the redshifting of the z filter
response.  In the HSTz images for $z<1.2$ many of the IDs are clearly
spiral or disk galaxies. This result is consistent with most of
  the galaxies in the sub-sample being due to star-formation. Beyond
$z\sim1.2$, as the z filter is redshifted further to the blue
and beyond the 4000\AA\ break , the galaxy morphologies become
less clear.  Mostly the objects appear to be an extension of the upper
end of the $z<1$ sample up to  $z\sim1.2$ but the nature of the
$z>1.2$ objects becomes less clear based on this analysis. 
  However, \citet{b15} shows that for this same, full sample, 
except for the
  the most luminous X-ray sources ($L_X > 5\times10^{43}$ ergs
  s$^{-1}$), the radio sources fit the radio-FIR relation. This result
  suggests that the typical radio source in our full radio 
sample, out to $z\sim 4$, is due to star-formation.

Twenty-two radio sources with sizes $> 1$\arcsec\ are
classified as binaries (or multiple systems). These systems have
similar radio and optical sizes with medians $\sim 20$kpc. Of course
their definition is largely dependent on the survey resolution, since
much smaller objects would be unresolved and much larger objects would
be cataloged as individual galaxies. Their median redshift is near
$z=1$ and their median $log(L_{20})=23.26$. Thus they appear
unremarkable except that they contribute about 10\% to the resolved
radio source sample.

\subsubsection{Radio versus HST Orientations}

\begin{figure}[htb!]
\includegraphics[width=1.03\columnwidth]{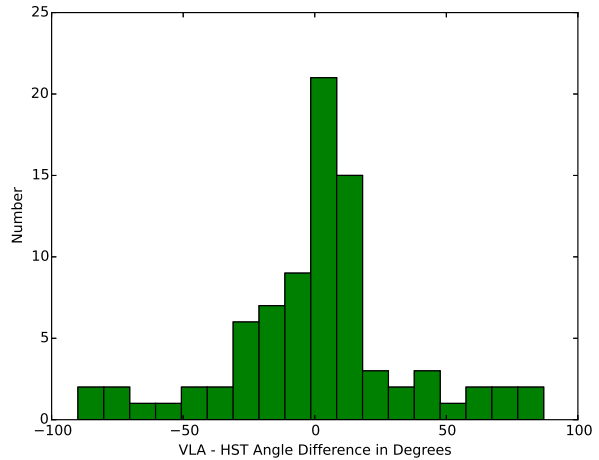}
\caption{Difference in orientation angles  for VLA radio sources with
  deconvolved sizes $>1\arcsec$ identified
  with individual galaxies versus orientation of the identifications
  from HSTz images. Only radio sources with orientation errors less than 20
  degrees are included.
\label{gorient}}
\end{figure}

It is also important to compare the orientations of the radio
emission and optical isophotes. For sources sizes $> 1$\arcsec,
we remeasured the position angle of the radio emission using JMFIT in
order to obtain the best estimate of the position angle. Sometimes the
smallest position angle error was obtained using an image with a
slightly different resolution than was used in the main catalog. We
also measured the optical position angles for the corresponding
optical identifications for the outer visible isophotes on HST
images. For the HST images, we used the z-band. The position angles and
sizes were measured down to the faintest isophote easily visible above the
noise. This scale tends to be larger than what one often picks as the
extent of the disk but is well defined on the images and mostly aligns
with any smaller scale disk.  Typically we estimate that we measured
the optical position angle to within 10 degrees. These results are
summarized in table 3.  For sources with JMFIT errors $< 20$ degrees,
we plot in figure~\ref{gorient} the position angle differences. 74\%
of the resulting sample are aligned within 30 degrees. In the majority
these cases, the identification is clearly a disk galaxy. Thus these
extended systems appear to have emission aligned with the disk, likely
from star-formation along the disk.

The remaining 24\%, apparently misaligned sources, show no outstanding
common property and are likely due to a combination of effects. Three
are very low S/N, $\le 5.3$, and based on the simulations could be
spurious resolved sources. Three more have S/N between 6.2 and 6.7.
The median S/N is 9.1, with one as high as 75.  Two appear to be
associated with E-galaxies and thus are AGN candidates.  One is
aligned with an inner structure but not with outer spiral arms. Most
appear to be disk galaxies, without obvious alignment with any
features in the image.

\begin{figure}[htb!]
\includegraphics[width=1.03\columnwidth]{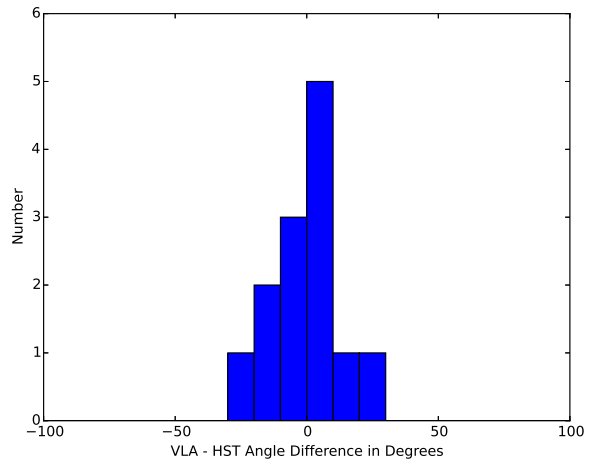}
\caption{Difference in orientation angles for individual VLA sources identified
  with multiple galaxies (mostly binaries) and with radio sizes
  $>1\arcsec$ in fitted size versus orientation of the identifications
  from HSTz images. Only radio sources with orientation errors less than 20
  degrees are included.
\label{borient}}
\end{figure}

In figure~\ref{borient}, we show a similar plot for radio sources
identified with binary galaxies (or multiple systems, mostly
binaries). Here the alignment is even better.  In addition to these
close binaries, there is a small sample of larger-scale blends
discussed in the previous section.  For other sources $>1$\arcsec,
there is some reason which prevents us from making a radio-optical
comparison. Either the optical identification is too small or the
radio or optical object is too round to measure an orientation
accurately. There is a tendency for the galaxies  imaged with
  HST z-band in these multiple systems to be smaller than the IDs
with single galaxies. The median isophotal size for the larger
  of the two optical galaxies in binaries is only $\sim 10$kpc. These
smaller galaxies in binaries also appear usually to have dust
obscuration in the nuclear region. 7/8 of the systems with its larger
galaxy $>15$ kpc have $z<1$ and 6/8 have $\log(L_{20}) < 23$. For the
14 smaller systems 13/14 have $z>1$ and $\log(L_{20}) > 23$.

\begin{figure}[htb!]
\includegraphics[width=1.03\columnwidth]{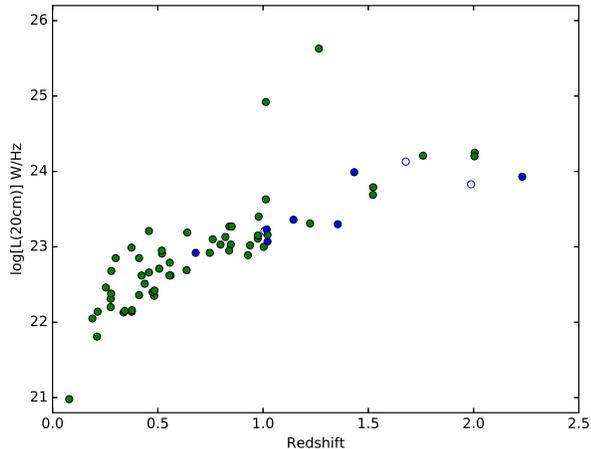}
\caption{Luminosity versus redshift for VLA sources $>1\arcsec$
  identified with individual galaxies (green circles) and binary
  systems (blue circles). Filled circles have spectroscopic redshifts,
  open circles have photometric redshift. Only radio sources with
  orientation errors less than 20 degrees and absolute Radio-HST
  position angle difference less than 30 degrees are plotted.  The
    LIRG/ULIRG boundary is $log(L_{20})\sim23.25$.
\label{anglelum}}
\end{figure}

In figure~\ref{anglelum}, the radio luminosity versus redshift is
plotted for galaxies and binaries with which have radio position angle
errors from AIPS JMFIT $< 20$\arcdeg\ and which are aligned to within
better than 30\arcdeg. As can be seen in the plot the aligned sample
extends well beyond the LIRG luminosity cutoff into the ULIRG class
and generally has typical luminosities near the LIRG/ULIRG luminosity
boundary.

 The conclusion is that, at least for $z \ltw 1$, the radio sources in the
survey with structure $>1$\arcsec\ tend to be
either large galaxies in the LIRG/ULIRG class with disk emission on scales
outside their nuclear regions likely driven by star-formation, or multiple
galaxy systems. 

\section{Discussion}

The median redshift for sources with measured redshifts for the
survey ($z=1.01$) marks a useful dividing point for understanding the
properties of the sources. For $z<1$, 68\% of the sources have
$\log(L_{20})$ between 22 and 23. 8\% have radio luminosities below
this range and 24\% above. For star-forming galaxies this range
corresponds roughly to LIRGs. As discussed earlier, locally it is
  well known that at $\log(L_{20}) < 23$, the radio luminosity from
  galaxies is dominated by star-formation \citep[e.g.][]{c02}.  Thus
76\% of the lower redshift half of the survey have luminosities in the
  range locally dominated by star-formation and most of those are in
  the range occupied by strongly star-forming galaxies but not
objects in the ULIRG class. Examination of the HST images reveals
many, perhaps most, of these objects appear to be dusty disk
galaxies, very consistent with strong star formation. For those
that have high enough S/N so that a size and position angle can be
measured, the radio emission is statistically aligned with the optical
major axis (figure~\ref{gorient}). Figure~\ref{size} suggests that at
the bottom of the catalog, we are likely underestimating the flux
densities of such galaxies and missing a significant number of them
since we are resolving them and cannot measure a statistically
significant size.  However, the ones we can analyze suggest that 
much of the disk is emitting star-formation-driven radio
emission. Figure~\ref{anglelum} shows that most of the radio-optical
aligned galaxies we have found have $z\ltw1$.

Above $z=1$, only about 6\% of the sources have $\log(L_{20}) < 23$.
The median redshift is 1.7. The median $\log(L_{20})$ is 23.72,
corresponding to $\sim 300$ M$\odot$/yr. 32\% have $\log(L_{20})> 24$
suggesting either SFR $> 500$ M$\odot$/yr, AGN-related emission, a
combination of AGN/star-formation or perhaps some other physics is
involved.  A few very large, luminous sources are clearly AGN, based on
their radio structure, but the rest need other information to
understand them, e.g. \citet{b15}. Furthermore \cite{b17} find that for
  $z>0.8$, $\sim40$\% of the radio sources in our GOODS-N sample with
  $\log(L_{20})>24$ are consistent with star-formation, based on
  comparison of the radio and submillimeter flux densities. The rest
  show some evidence of an AGN contribution. Especially for
  $\log(L_{20})>24.5$, all the sources appear to be AGN suggesting a
  cutoff in the maximum SFR at $\sim500$ M$\odot$/yr. 

Besides the ``standard'' radio population there clearly are some other
origins of sources in the table. Small subsets are either binary
galaxies with both members emitting or blends of emission from two or
more apparently unrelated galaxies. There also is a small population
of sources without identifications. Objects so faint seem likely to be
very high z and/or very dusty galaxies. If AGN-driven bubble
  sources or star-forming radio galaxies dominated by wind-related
  emission were important, we would have expected a significant number
  of diffuse sources not aligned with the optical galaxy axis and
  often larger than the optical galaxy image. We do not find such
  populations. It remains possible that they remain hidden due to the
  star-formation-driven or AGN-driven emission dominating the regions
  imaged. At the sensitivity level of this survey, star-formation and
  some AGN activity, especially at higher redshifts, seem to dominate
  the observed radio emission.

 The most surprising result is the statistical alignment of the sources
  $>1$\arcsec\ in size with the extended optical disks, especially with $z<1$.
  This implies that there is a significant population of galaxies in the
  LIRG/ULIRG class which are not dominated by nuclear star-formation but
  have most of their star-forming radio emission extended along the galaxy
  disk, typically $\sim 10$kpc. What fraction of all the radio galaxies in the
  survey have this extended star-forming emission is hard to pin down because
  for the weaker cataloged sources the S/N, even given the convolved searches,
  would not be adequate to measure the extensions at significant levels. Deeper
  surveys are needed to study this problem.

This division into two redshift ranges does not suggest necessarily
that there is a fundamental change in the radio source populations at
$z=1$, only that the survey selection effects produce such a
division. This survey is not deep enough to study the LIRG population
much beyond $z=1$. The resolution and surface brightness sensitivity
is just high enough to to resolve a significant fraction of the
sources. Since locally such sources show extended winds which emit in
the radio with steep radio spectral indices, higher rest-frame
frequencies may miss that part of the story, especially for higher
redshifts.  Combined with the limited surface brightness
  sensitivity even at 1.5GHz we may be missing a lot at the higher
  rest frequencies especially for $z>1$. Deep surveys at lower
  frequencies would be a useful check.

\section{Conclusions}

A new radio catalog based on a deep, 1525 MHz VLA radio image
($2.2\mu$Jy {\it rms}  near the field center of the
1.6\arcsec\ image) is reported, along with its properties, including
optical/NIR identifications. 795 sources were detected within
9\arcmin\ of the nominal GOODS-N center position. With modern
  ground-based NIR imaging depths and with HST optical imaging, the
  vast majority of radio sources with at our radio limits have been
  identified reliably with $<2$\% of the sources remaining
  unidentified or misidentified. Especially for $z<1$, a large
subset of the survey has radio sizes $> 1$\arcsec\ .  The size and
alignment of the radio emission for those sources with the HST images
of disk galaxies suggests that there is a large population of
LIRG/ULIRGs,  with most of their star-formation not isolated in the
  galaxy nucleus, but with strong star-formation extending
along the 10kpc-scale disk. This survey naturally divides at the
  median redshift near $z=1$, into 1) a $z<1$ population mostly very
  consistent with strongly star-forming disk galaxies and many with
  resolved star-forming regions and 2) a $z>1$ sample typically
  with much higher SFRs and/or AGN emission.

\section{Acknowledgments}

This research has made use of the VizieR catalogue access tool, CDS,
Strasbourg, France. The original description of the VizieR service was
published in A\&AS 143, 23.

 The HST GOODS-N cutout images presented in this paper were
  obtained from the Mikulski Archive for Space Telescopes
  (MAST). STScI is operated by the Association of Universities for
  Research in Astronomy, Inc., under NASA contract NAS5-26555.
  Support for MAST for non-HST data is provided by the NASA Office of
  Space Science via grant NNX09AF08G and by other grants and
  contracts.

The author thanks Bill Cotton, Emanuele Daddi, Jean Eilek
and Ken Kellermann for comments on the text.

\clearpage

\end{document}